\documentstyle[epsfig,prl,aps]{revtex}
\begin{document}
\draft
\twocolumn[\hsize\textwidth\columnwidth\hsize\csname
@twocolumnfalse\endcsname
\title{ VARIABLE COSMOLOGICAL CONSTANT - GEOMETRY AND PHYSICS}
\author{Irina Dymnikova}
\address{Institute of Mathematics, Informatics and Physics,\\
         University of Warmia and Mazury in Olsztyn,
         10-561 Olsztyn, Poland}
\maketitle

\vskip0.2in
{\bf The talk at the International Seminar "Mathematical Physics and
Applied Mathematics" in memory of Professor Georgij Abramovich Grinberg 
(1900-1991), 16 June 2000, St-Petersburg, Russia.}
\vskip0.1in
{\it DEDICATION

The results presented here probably would have never 
been obtained without unforgettable hours spent by author,
throughout twenty years 1968-1988, in conversations with people 
from the Mathematical Physics Department of the A.F. Ioffe Institute 
in Leningrad. It was oasis of deep respect for human being 
and products of his thinking. People of this Department created atmosphere
in which one immediately got a feeling that not only he is able to produce 
something but this is just his natural state. Coming there 
I, for one, felt like a fish moved from a pan into a water. 
I thank the Providence for this gift and dedicate this paper 
to the blessed memory of Georgij Abramovich Grinberg, 
Nikolaj Nikolajevich Lebedev and Jakov Solomonovich Ufland.}

\begin{abstract} 
{\bf Abstract}

We describe the dynamics of a cosmological
term in the spherically symmetric case by an $r$-dependent second rank 
symmetric 
tensor $\Lambda_{\mu\nu}$ invariant under boosts in the radial direction. 
The cosmological tensor $\Lambda_{\mu\nu}$ represents the extension of 
the Einstein cosmological term  $\Lambda g_{\mu\nu}$
which allows it to be variable. 
This possibility is based on the Petrov classification scheme and Einstein
field equations in the spherically symmetric case.
The inflationary equation of state $p=-\rho$ is satisfied by the radial
pressure, ${p}_r^{\Lambda}=-{\rho}^{\Lambda}$.
The tangential pressure ${p}_{\perp}^{\Lambda}$ is calculated from the 
conservation equation $\Lambda^{\mu}_{\nu;\mu}=0$.
The solutions of Einstein equations with cosmological
term  $\Lambda_{\mu\nu}$ describe
several types of globally regular self-gravitating vacuum configurations
including vacuum nonsingular black holes. In this case global structure
of space-time contains an infinite set of black and white holes,
whose past and future singularities are replaced with the value 
of cosmological constant $\Lambda$ 
of the scale of symmetry restoration, at the background of
asymptotically flat or asymptotically de Sitter (with $\lambda<\Lambda$)
universes. Geodesics structure of space-time demonstrates 
possibility of travelling into other universes through the interior
of a $\Lambda$ black hole.
In the course of Hawking evaporation of a $\Lambda$BH, a second-order 
phase transition occurs, and globally regular configuration evolves 
towards a self-gravitating particlelike structure
($\Lambda$ particle) at the background of Minkowski or de Sitter space.
\end{abstract}
\vspace{0.2cm}
]

\section{Introduction}

The cosmological constant was 
introduced by Einstein in 1917 as the universal repulsion to make 
the Universe static in accordance with generally accepted picture
of that time. 
The Einstein equations with a cosmological term read 
$$
   G_{\mu\nu}+\Lambda g_{\mu\nu}=-\kappa T_{\mu\nu}\eqno(1)$$
where $G_{\mu\nu}$ is the Einstein tensor, $g_{\mu\nu}$ is the metric, 
$T_{\mu\nu}$ is the stress-energy tensor
of a matter, $\kappa$ is the Einstein gravitational constant
related to the Newton gravitational constant $G$ by
$\kappa =8\pi G c^{-4}$, and $\Lambda$ is the cosmological constant.
In the absence of  matter described by $T_{\mu\nu}$, $\Lambda$
must be constant, since the Bianchi identities guarantee vanishing
covariant divergence of the Einstein tensor, $G^{\mu\nu}_{~~;\nu}=0$,
while $g^{\mu\nu}_{~~;\nu}=0$ by definition.

In 1922 Friedmann found nonstationary solutions to the Einstein
equations (1) describing an expanding universe,
and in 1926-1929 the Hubble observations established the
fact that our Universe is expanding.

At the late 40-s cosmological constant was called for second time
to make the Universe stationary. In steady-state Bondi-Gold-Hoyle
cosmology \cite{bg,hoyle}
permanent creation of matter has been assumed, with cosmological constant
related to $C$-field responsible for this process. 
\vskip0.1in

Third coming of cosmological constant in the 70-s was quite spectacular.
It was called for to make the very early Universe  blowing up.
The main reason came from the problem of initial conditions 
for an expanding Universe. 
A closed Universe starting from initial Planckian density
$\rho_{Pl}\sim 10^{94}g~cm^{-3}$ and evolving in accordance with
the standard FRW cosmology,
would recollapse back in Planckian time $t_{age}\sim 10^{-43}s$.
For open or spatially flat Universe we would have today density
$\rho< 10^{-90}g~cm^{-3}$. Looking back we would have to conclude that 
observable Universe would expand from a region of $a(t_{Pl})\sim 10^{-13}cm$
that contained a huge number of causally disconnected regions of
Planckian size, which makes impossible to explain presently observed
homogeneity and isotropy of the Universe (for review see \cite{kolb}).

Analyzing in 1930 the problem of 
the Universe evolution 
Sir A. Eddington made two remarks \cite{ed}. First concerned the problem
of initial conditions:  
"Difficulty in admitting this situation is that
it seems to require some sudden and incomprehensible origin of all
things". Second concerned the de Sitter solution.
In the case of absence of a matter described by $T_{\mu\nu}$,
the solution to the Einstein equation (1), found by de Sitter in 1917,
is de Sitter geometry with constant positive curvature $R=4\Lambda$
$$
   ds^2=\biggl(1-\frac{\Lambda r^2}{3}\biggr)dt^2
   -\biggl(1-\frac{\Lambda r^2}{3}\biggr)^{-1}dr^2-r^2 d\Omega^2,$$
where $d\Omega^2$ is the line element on the unit sphere.
Main feature of this geometry is divergence of geodesics 
- in de Sitter geometry gravity effectively acts as a repulsion. 
Having this in mind, Eddington asked:
"May be this is why the Universe is expanding?"

During several next decades, the physical essence
of the de Sitter solution remained obscure. In modern physics
it has been mainly used as a simple testing ground for developing
the quantum field technics in a curved space-time.
In 1965 Gliner understood that de Sitter geometry is generated
by a vacuum with nonzero energy density\cite{gliner}.
Shifting a cosmological term $\Lambda g_{\mu\nu}$ 
from the left-hand side  to the right-hand side of Einstein equation (1)
corresponds to introducing a stress-energy tensor
$$
   T_{\mu\nu}^{vac} =(\kappa)^{-1}\Lambda g_{\mu\nu} =
    \rho_{vac} g_{\mu\nu};~~~\rho_{vac}= (\kappa)^{-1}\Lambda\eqno(2)$$
whose equation of state is
$$
   p = -\rho_{vac}\eqno(3)$$
This stress-energy tensor has an infinite set of comoving 
reference frames.  
An observer moving through the medium with the stress-energy
tensor (2), cannot in principle measure his velocity with respect to it, 
since his comoving reference frame is also comoving for 
$T_{\mu\nu}^{vac}$ \cite{gliner}.

In 1970 Gliner suggested that de Sitter vacuum can be
initial state for an expanding Universe \cite{gl70}. In 1975 
the first nonsingular cosmological model was proposed 
with the initial de Sitter stage \cite{us75}.
We have shown that the transition from $\Lambda$ dominated to the
radiation dominated stage produces the growth of the scale factor
by about 30 orders of magnitude for the Planck scale vacuum, 
accompanied by needed growth in the entropy -
the result confirmed by all inflationary models involving various
mechanisms responsible for $\Lambda$ (for review see \cite{olive}).   

Cosmological constant or huge initial vacuum energy density
is crucially needed to guarantee surviving of the Universe to its present
size and density as well as to explain its observable homogeneity and
isotropy \cite{gl70,us75,star,guth,lin,as}. Inflationary expansion
is supposed to begin at the Grand Unification scale of symmetry breaking 
$E_{GUT}\sim{10^{15}}$ GeV, blowing up a small causally connected
region to a size sufficient to explain puzzles of standard Big Bang
cosmology (for review and list of puzzles see, e.g., \cite{kolb})
-although leaving without explanation the puzzle of cosmological
constant itself:
{\it If it was so huge then - how it became so small now?} 

\vskip0.1in

{\bf The main puzzle of cosmological constant -} 

In Quantum Field Theory (QFT), the vacuum stress-energy
tensor has the form \cite{DEW,adler,weinberg} 
$$
   <T_{\mu\nu}>=<\rho_{vac}>g_{\mu\nu}$$ 
which behaves like an effective cosmological term 
with $\Lambda=\kappa <\rho_{vac}>$. In QFT a relativistic field 
is considered as a collection of harmonic oscillators 
of all possible frequencies.
Vacuum in QFT is a superposition of ground states of all fields.
Therefore it has the energy
$$
   E_{vac}=\sum{{1\over2}\hbar\omega},\eqno(4)$$
where ${1\over2}\hbar\omega$ is a zero-point energy 
of each particular field mode.
The expression (4) follows formally from commutation relations 
which in turn follow from the uncertainty principle applied 
to a picture of a filed  as a linear superposition of oscillators. 
Even for a single harmonic oscillator the uncertainty principle 
does not allow a particle to be fixed in a state
with fixed zero kinetic and zero potential energy. Therefore, 
for quantum mechanical reasons, zero-point vacuum energy cannot
be put equal zero. For aesthetic reasons it could be removed by some
subtracting procedure (normal ordering), although it is called back each time
when one has to calculate physical effects due to vacuum polarization
- for example, Casimir
effect \cite{cas} which is quite measurable \cite{spa}.

An upper cutoff for a vacuum energy density is estimated 
by an energy scale at which our
confidence in the formalism of QFT is broken. It is widely believed that
it is the Planckian energy $E_{Pl}\sim{10^{19} GeV}\sim{10^{16} erg}$
that marks a point where QFT breaks down due to
quantum gravity effects (see, e.g., \cite{kraus}). It gives 
$$
   <\rho_{vac}>\sim{10^{94}g~ cm^{-3}}\eqno(5)$$
According to General Relativity, a vacuum contributes to gravity by Einstein
equations (1). Therefore gravitational influence of a vacuum is not 
avoidable in principle.
With vacuum density (5) its manifestations would be quite dramatic.
For example, the cosmic microwave background radiation would have cooled
below $3 K$ in the first $10^{-41}s$ after Big Bang, and expansion rate
(Hubble parameter) would be about a factor of $10^{61}$ larger than that
observed today \cite{kolb}.

This is the main puzzle of the cosmological constant. According to QFT it must
be huge, but according to observations it is for sure very small now.
Typically in QFT vacuum energy density is postulated
to be zero, in the hope that in future good theory there will be found some
cancellation mechanism which zeros out its ultimate value, or 
that quantum-cosmological considerations will 
favor a zero value of cosmological constant \cite{kraus}.

On the other hand, from the cosmological point of view, cosmological 
constant can not be put equal
zero everywhere and forever, since inflation demands it to be 
huge at the beginning of the Universe
evolution, while astronomical data give evidence \cite{kraus,ostriker,bahcall} 
that cosmological constant today is comparable with the average density
in the Universe 
$$
   \rho_{today}\sim{10^{-30}g~cm^{-3}}\eqno(6)$$

\vskip0.1in

{\bf Observational case for a cosmological constant -}
The key cosmological parameter to decide if cosmological constant  
is zero or not today, is the product of the Hubble parameter and the age 
of the Universe $H_0t_0$ (index zero labels values today).
For models without cosmological constant it never exceeds 1. 
In the presence of a nonzero cosmological constant it is possible 
for the Universe to have this product exceeding 1 (see, e.g.,\cite{kolb}).
If Hubble parameter and age of the Universe as measured from high redshift
would be found to satisfy the bound $H_0t_0 > 1$, it would require a term
in the expansion rate equation that acts as a cosmological constant. 
Therefore the definitive measurement of $H_0t_0 > 1$ would necessitate 
a non-zero cosmological constant today or the abandonment of 
standard big bang cosmology \cite{kraus}.

The most pressing piece of data in favour of nonzero cosmological
constant  involves the estimates of the age of 
the Universe as compared with the estimates of the Hubble parameter.
With taking into account uncertainties in models
the best fit to guarantee consensus between all observational constraints
is \cite{kraus,ostriker,bahcall}
$$
   H_0=(70-80) km~s^{-1}Mpc^{-1},~~~t_0=(13-16)\pm 3 Gy,$$ 
   $$ \Omega_{matter}=0.3-0.4,~~~ \Omega_{\Lambda}=0.6-0.7,$$
where $\Omega=\rho_{today}/\rho_{cr}$, and the critical density
$\rho_{cr}$, which correspond to $\Omega=1$, is given by (6).

Confrontation of models with observations in cosmology 
as well as the inflationary paradigm, compellingly favour  
treating the cosmological constant  
as a variable dynamical quantity .

\vskip0.1in

{\bf $\Lambda$ variability -} The idea that $\Lambda$ might be variable 
has been studied for more
than two decades (see \cite{adler,weinberg} and references therein).
In a recent paper on $\Lambda$-variability,
Overduin and Cooperstock distinguish three approaches \cite{overduin}. 
In the first approach
$\Lambda g_{\mu\nu}$ is shifted onto the right-hand side of the field
equations (1) and treated as  part of the matter content. This
approach, originated mainly from Soviet physics school and
characterized by Overduin and Cooperstock as 
connected to dialectic materialism,
goes back to Gliner who interpreted $\Lambda g_{\mu\nu}$ 
as a vacuum stress-energy tensor, to Zel'dovich 
who connected $\Lambda$ with the gravitational interaction of virtual 
particles \cite{zeldovich}, and to Linde who suggested that $\Lambda$ 
can vary \cite{andrei}. 

In contrast, idealistic approach prefers 
to keep $\Lambda$ on the left-hand side of the Eq.(1) and treat it as a
constant of nature. The third approach, allowing
$\Lambda$ to vary while keeping it on the left-hand side as a geometrical
entity, was first applied by Dolgov in a model in which
a classically unstable scalar field, non-minimally coupled to gravity,
develops a negative energy density cancelling the initial positive
value of a cosmological constant $\Lambda$ \cite{dolgov}.
 
Whenever variability of $\Lambda$ is possible, it requires
the presence of some matter source other than
$T_{\mu\nu}=(\kappa)^{-1}\Lambda g_{\mu\nu}$, since the conservation
equation $G^{\mu\nu}_{~~;\nu}=0$ implies $\Lambda$=const in this case.
This requirement makes it impossible
to introduce a cosmological term as variable in itself. 
However, it is possible for a stress-energy tensor
other than $\Lambda g_{\mu\nu}$.

In the spherically symmetric case such a possibility
is suggested by the Petrov 
classification scheme \cite{petrov} and by the Einstein field 
equations. The algebraic structure of the stress-energy tensor in this case
is $T_r^r=T_t^t;~T_{\theta}^{\theta}=T_{\phi}^{\phi}$ \cite{me92}.
This tensor is invariant under boosts in the radial direction,
and describes  a spherically symmetric vacuum which 
represents the extension of the Einstein cosmological 
term $\Lambda g_{\mu\nu}$
to the spherically symmetric $r-$dependent cosmological tensor
$\Lambda_{\mu\nu}$ \cite{lambda}.

\vskip0.1in

{\bf Nonsingular black hole -} The original motivation for such 
an extension was to replace a black hole singularity by de Sitter 
vacuum core.
The idea goes back to the 1965 papers by Sakharov who considered
the equation of state $p=-\rho$ as arising at the superhigh
densities \cite{sakharov}, and by Gliner who suggested
that de Sitter vacuum could be a final state in the
gravitational collapse \cite{gliner}. 

The papers developing this idea
appeared in the  80-s as inspired by successes of inflationary
paradigm. In those papers direct matching of
Schwarzschild metric to the de Sitter metric was considered 
using thin shell approach \cite{bern,fg,shen}. In this approach 
the equation of state changes from
$p=0,~ ~ \rho=0$ outside to $p=-\rho_{Pl}\sim{10^{94}g~cm^{-3}}$ within
a junction layer of Planckian thickness $\delta\sim{l_{Pl}}
\sim{10^{-33}cm}$, and the resulting metric has a jump at the junction
surface. The situation was analyzed by Poisson and Israel
who suggested to insert a layer of a noninflationary material of the
uncertain depth at the interface, in which 
a geometry remains effectively 
classical and governed by the Einstein equations with a source term 
representing vacuum polarization effects \cite{werner}.

The exact analytic solution describing de Sitter-Schwarzschild transition 
in general case of a distributed density profile, has been 
found in the Ref \cite{me92} in the frame of a simple semiclassical model for  
a density profile due to vacuum polarization in a spherically symmetric 
gravitational field \cite{me96}.

In the case of de Sitter-Schwarzschild transition the variable
cosmological term $\Lambda_{\mu\nu}$  connects in 
a smooth way two vacuum states: de Sitter vacuum 
$T_{\mu\nu}=(\kappa)^{-1}\Lambda g_{\mu\nu}$ replacing a singularity
at the origin, and Minkowski vacuum $T_{\mu\nu}=0$ at infinity.
For any density profile satisfying the conditions of finiteness of a mass
and of a density, the solution to the Einstein equations
 describes  the  globally regular de Sitter-Schwarzschild geometry,
asymptotically Schwarzschild as $r\rightarrow{\infty}$ and
asymptotically de Sitter as $r\rightarrow 0$ \cite{lambda,me96,me99}.
In the range of masses 
$M\geq M_{crit}\simeq{0.3M_{Pl}\sqrt{\Lambda_{Pl}/\Lambda}}$ 
this geometry represents a $\Lambda$ black hole ($\Lambda$BH),
while for $M<M_{cr}$ - a $\Lambda$ particle, self-gravitating
particlelike structure made up of self-gravitating
 spherically symmetric vacuum $\Lambda_{\mu\nu}$ \cite{me96,lambda}. 
In the course of Hawking evaporation of a $\Lambda$BH, a second-order 
phase transition occurs, and globally regular configuration evolves 
towards a cold remnant which is self-gravitating particlelike structure
($\Lambda$ particle) at the background of Minkowski or de Sitter space.

In the black hole case the global structure of space-time shows that
in place of a former singularity there arises an infinite chain
of structures including $\Lambda$ white holes ($\Lambda$WH) and asymptotically
flat universes. 

\vskip0.1in

{\bf Universes inside a black hole -}
The idea of a baby Universe inside a black hole 
has been proposed by Farhi and Guth (FG) in 1987 \cite{fg} 
as the idea of creation of a universe 
in the laboratory starting from a false vacuum bubble in the Minkowski
space. 
FG studied an expanding spherical de Sitter bubble separated by thin
wall from the outside region where the geometry is Schwarzschild.
The global structure of space-time in this case 
implies that an expanding bubble must be associated
with an initial spacelike singularity.
The initial singularity clearly represents
a singular initial value configuration. Therefore Farhi and Guth 
concluded that the requirement of initial singularity would be
an obstacle to the creation of a universe inside a black hole \cite{fg}. 

In 1990
Farhi, Guth and Guven (FGG)
studied the model in which the initial bubble is small enough
to be produced without initial singularity \cite{fgg}.
A small bubble classically could not become a universe - instead it would
reach a maximum radius and then collapse. FGG
investigated the possibility that quantum effects allow the bubble
to tunnel into the larger bubble of the same mass which
for an external observer disappears
beyond the black hole horizon, whereas on the inside the bubble would
classically evolve to become a new universe \cite{fgg}.
 
Arising a new universe inside a black hole has been 
considered also by Frolov, Markov, and Mukhanov (FMM) 
in 1989-1990 \cite{valera}. 
The difference of FMM from FG approach is that the FG assumption - 
the existence of a global Cauchy surface - has been violated, due to
the existence of the Cauchy 
horizon, firstly noticed by Poisson and Israel in 1988 \cite{werner},
which implies the absence of a global Cauchy surface. 

Both FG and FMM models are based on matching Schwarzschild metric
outside to de Sitter metric inside a junction layer of
Planckian thickness $~ l_{Pl}\sim{10^{-33}}$cm.
The case of direct matching clearly corresponds to arising
of a closed or semiclosed world inside a black hole \cite{valera}.

In the case of a distributed density profile - a $\Lambda$BH -
the dynamical situation is different. 
The global structure of space-time implies the possibility
of existence of an infinite numbers of 
vacuum-dominated universes inside a $\Lambda$BH, 
and geodesics structure implies possibility of travelling to them. 
De Sitter vacuum exists near the surface of former singularity $r=0$.  
This region is the part of a $\Lambda$WH, where, as a result of a white
hole or de Sitter instability, there is possible quantum creation
of  baby universes which are disjoint
from each other and which can be not only closed, but also open
and flat \cite{us99}.

\vskip0.1in

{\bf Particlelike structure - } In the course of the
Hawking evaporation of a $\Lambda$BH a second-order phase 
transition occurs, and a globally regular configuration 
evolves toward a vacuum-made particlelike 
structure described by de Sitter-Schwarzschild geometry
for the range of masses $M<M_{crit}$ \cite{me96}. 
Such a configuration can be applied to estimate a lower
limit on sizes of fundamental particles (particles which do not display 
a substructure) \cite{zur}.  

In QFT particles are treated as point-like. Experiments give the upper
limits on  sizes of fundamental particles (FP)
defined by a characteristic size of the region of interaction
in the relevant reaction (see review in \cite{zur}). 
One can expect that the lower limits on particle sizes 
are determined by gravitational interaction \cite{zur}.
In the case of a FP its gravitational size cannot be defined by the
Schwarzschild gravitational radius $r_g=2GMc^{-2}$. 
Quantum mechanics constrains any size
from below by the Planck length $l_{Pl}\sim{10^{-33}}$ cm, and for any
elementary particle its Schwarzschild radius is many orders of magnitude
smaller than $l_{Pl}$. 

The Schwarzschild gravitational radius $r_g$ comes from the Schwarzschild
solution which implies a point-like mass and is singular at $r=0$.
De Sitter-Schwarzschild geometry represents the nonsingular modification
of the Schwarzschild geometry, and depends on the limiting vacuum density
$\rho_{vac}$ at $r=0$. For $M<M_{cr}$ it describes a neutral self-gravitating
particlelike structure with de Sitter vacuum core related to its
gravitational mass \cite{me96,lambda}.
 Now mass is not point-like but distributed,
and most of it is within  the characteristic size
$\sim{(m/\rho_{vac})^{1/3}}$, where  $\rho_{vac}$ is the density
at  $r=0$ \cite{me96,haifa}.

In the context of spontaneous symmetry breaking $\rho_{vac}$
is related to the potential of a scalar field in its symmetric
state (false vacuum). In the context of the Einstein-Yang-Mills-Higgs
(EYMH) self-gravitating non-Abelian structures including black holes,
$\rho_{vac}$ is related to symmetry restoration in the origin.
In a neutral branch of EYMH black hole solutions
a non-Abelian structure can be approximated as a sphere of
a uniform vacuum density $\rho_{vac}$ whose radius is the Compton
wavelength of the massive non-Abelian field (see \cite{maeda} and
references therein).

Results obtained in the frame of both EYMH systems and de Sitter-Schwarzschild
geometry suggest that a mass of a FP can be related to a gravitationally
induced core with the de Sitter vacuum at $r=0$. We can assume thus
that whatever would be a mechanism of a mass generation,
a FP must have an internal vacuum core related to its mass and a finite size
defined by gravity.
With this assumption we are able to set the lower limits on FP sizes
by sizes of their vacuum cores as defined by de Sitter-Schwarzschild geometry.
It gives us also an upper limit on the mass of a Higgs scalar \cite{zur}.

\vskip0.1in

{\bf Structure of this paper -}
This paper summarizes our results on variable cosmological term
$\Lambda_{\mu\nu}$ in the spherically symmetric case. It is organized 
as follows. 
The r-dependent cosmological term  
$\Lambda_{\mu\nu}$ is introduced in Section II. In Section III we present 
de Sitter-Schwarzschild geometry including both black hole
sector and particlelike 
sector. In Section IV we outline our results on two-lambda
geometry which is the extension of de Sitter-Schwarzschild geometry
to the case of nonzero value of cosmological constant at infinity.

\section{Cosmological term $\Lambda_{\mu\nu}$}

In the spherically symmetric static case a line element can be
written in the form \cite{tolman}
$$
   ds^2 = e^{\mu(r)}dt^2 - e^{\nu(r)} dr^2 - r^2 d\Omega^2\eqno(7)$$
where $d\Omega^2$ is the line element on the unit sphere.
The Einstein equations read
$$ 
   \kappa T_t^t = e^{-\nu}\biggl(\frac{{\nu}^{\prime}}{r}
   -\frac{1}{r^2}\biggr)+\frac{1}{r^2}\eqno(8)$$
$$
   \kappa T_r^r = -e^{-\nu} \biggl(\frac{{\mu}^{\prime}}{r}
   +\frac{1}{r^2}\biggr)+\frac{1}{r^2}\eqno(9)$$
$$
   \kappa T_{\theta}^{\theta}=8\pi G T_{\phi}^{\phi}=$$
     $$-e^{-\nu}\biggl(\frac{{\mu}^{\prime\prime}}{2}
    +\frac{{{\mu}^{\prime}}^2}{4}+\frac{{\mu}^{\prime}
     -{\nu}^{\prime}}{2r}-\frac{{\mu}^{\prime}
      {\nu}^{\prime}}{4}\biggr)\eqno(10)$$
A prime denotes differentiation with respect to $r$.
In the case of 
$$
   T_{\mu\nu}={\kappa}^{-1} \Lambda g_{\mu\nu} 
        = \rho_{vac} g_{\mu\nu}\eqno(11)$$
the solution is the de Sitter geometry with constant positive curvature
$R=4\Lambda$. The line element is
$$
   ds^2=\biggl(1-\frac{\Lambda r^2}{3}\biggr)dt^2-
   \biggl({1-\frac{\Lambda r^2}{3}}\biggr)^{-1}dr^2-r^2 d\Omega^2\eqno(12)$$
There is the causal horizon in this geometry defined by
$$
   r_0^2=\frac{3}{\Lambda}=\frac{3}{\kappa\rho_{vac}}\eqno(13)$$

The algebraic structure of the stress-energy tensor, corresponding to 
a cosmological term $\Lambda g_{\mu\nu}$, is
$$
   T_t^t=T_r^r=T_{\theta}^{\theta}=T_{\phi}^{\phi}; ~~~p=-\rho\eqno(14)$$

In the Petrov classification scheme \cite{petrov}
stress-energy tensors are
classified on the basis of their algebraic structure.
When the elementary divisors of the matrix 
$T_{\mu\nu}-\beta g_{\mu\nu}$
(i.e., the eigenvalues of $T_{\mu\nu}$) are real, the eigenvectors of
$T_{\mu\nu}$ are nonisotropic and form a comoving reference
frame. Its timelike vector represents a velocity. The classification
of the possible 
algebraic structures of stress-energy tensors satisfying the above
conditions contains five 
possible types: [IIII],[I(III)], [II(II)], [(II)(II)], [(IIII)].
 The first symbol denotes the eigenvalue related to the timelike eigenvector.
Parentheses combine equal (degenerate) eigenvalues.
A comoving reference frame is defined uniquely
 if and only if none of the
spacelike eigenvalues ${\beta}_{\alpha}(\alpha=1,2,3)$  coincides with
a timelike eigenvalue ${\beta}_0$. Otherwise there exists an infinite set
of comoving reference frames. 

In this scheme the de Sitter
stress-energy tensor (11) is represented by [(IIII)] (all eigenvalues 
being equal)
and classified as a vacuum tensor due to the absence of a preferred comoving
reference frame \cite{gliner}.
In the spherically symmetric case it is possible, by the
same definition, to introduce  an $r$-dependent vacuum
stress-energy tensor with the algebraic structure \cite{me92} 
$$
   T_t^t=T_r^r;~~T_{\theta}^{\theta}=T_{\phi}^{\phi}\eqno(15)$$
In the Petrov classification scheme this stress-energy tensor 
is denoted by [(II)(II)].
It has an infinite set
of comoving reference frames, since it is invariant under rotations
in the $(r,t)$ plane. Therefore an observer moving through it
cannot in principle measure the radial component of his velocity.  
The stress-energy tensor (15)
describes a spherically symmetric anisotropic vacuum 
invariant under the boosts in the radial direction \cite{me92}.

The conservation equation $T^{\mu\nu}_{~~;\nu}=0$ gives the $r$-dependent
equation of state \cite{werner,me92}
$$
   p_r=-\rho;~~p_{\perp}=p_r+\frac{r}{2}\frac{dp_r}{dr},\eqno(16)$$
where $\rho=T_t^t$ is the density, $p_r=-T_r^r$ is the radial
pressure, and $p_{\perp}=-T_{\theta}^{\theta}=-T_{\phi}^{\phi}$
is the tangential pressure. 
In this case  equations (7)-(8) reduce  to the equation
$$
   \kappa\rho=e^{-\nu}\biggl(\frac{{{\nu}^{\prime}}}{r}-
    \frac{1}{r^2}\biggr)+\frac{1}{r^2}\eqno(17)$$
whose solution is
$$
   g_{00}=e^{-\nu(r)}=1-\frac{2G{\cal M}(r)}{r};~~{\cal M}(r)
      =4\pi\int_0^r{\rho(x)x^2dx}\eqno(18)$$
and the line element is given by 
$$
   ds^2=\biggl(1-\frac{2G{\cal M}(r)}{r}\biggr)dt^2-
   \biggl(1-\frac{2G{\cal M}(r)}{r}\biggr)^{-1}dr^2-r^2d\Omega^2\eqno(19)$$
If we require the density $\rho(r)$ to vanish as $r\rightarrow
{\infty}$ quicker then $r^{-3}$, then the metric (19) for large $r$ 
has the Schwarzschild form 
$$
   ds^2=\biggl(1-\frac{2GM}{r}\biggr)dt^2-
     \biggl(1-\frac{2GM}{r}\biggr)^{-1}dr^2-r^2d\Omega^2\eqno(20)$$
with
$$
   M=4\pi\int_0^{\infty}{\rho(r)r^2dr} < \infty\eqno(21)$$
If we impose the boundary condition of de Sitter behaviour (12)
at $r\rightarrow 0$, the form of the mass function
${\cal M}(r)$ in the limit of small $r$ must be 
$$
   {\cal M}(r)=\frac{\Lambda}{6G}r^3=\frac{4\pi}{3}\rho_{vac}r^3\eqno(22)$$
For any density profile satisfying conditions (21)-(22),
the metric (19) describes a globally regular de Sitter-Schwarzschild geometry,
asymptotically Schwarzschild as $r\rightarrow{\infty}$ and
asymptotically de Sitter as $r\rightarrow 0$ \cite{me96,me99}.

The stress-energy tensor (15) responsible for this geometry 
connects in a smooth way two vacuum states:
de Sitter vacuum (11) at the origin and Minkowski vacuum $T_{\mu\nu}=0$
at infinity. The inflationary equation of state $p=-\rho$
remains valid for the radial component of a pressure.
This makes it possible to treat the stress-energy tensor (15) as
corresponding to  an $r$-dependent  cosmological term 
$\Lambda_{\mu\nu}$, varying from $\Lambda_{\mu\nu}=\Lambda g_{\mu\nu}$ 
as $r\rightarrow 0$ to $\Lambda_{\mu\nu}=0$ as $r\rightarrow\infty$, and 
satisfying the equation of state (16) with $\rho^{\Lambda}=\Lambda^t_t$,
$p_r^{\Lambda}=-\Lambda^r_r$ and 
$p_{\perp}^{\Lambda}=-\Lambda^{\theta}_{\theta}=-\Lambda^{\phi}_{\phi}$. 

If we modify the density profile to allow a non-zero value
of cosmological constant $\lambda$ as $r\rightarrow {\infty}$, putting
$$
    T_t^t(r)=\rho(r)+{\kappa}^{-1}\lambda,\eqno(23)$$
we obtain the metric \cite{us97}
$$
   ds^2=\biggl(1-\frac{2G{\cal M}(r)}{r}
    -\frac{\lambda r^2}{3}\biggr)dt^2- $$
   $$\biggl({1-\frac{2G{\cal M}(r)}{r}
   -\frac{\lambda r^2}{3}}\biggr)^{-1} dr^2-r^2d\Omega^2\eqno(24)$$
whose asymptotics are the de Sitter metric (12) with $\lambda$ as
$r\rightarrow{\infty}$ and with
$(\Lambda +\lambda)$ as $r\rightarrow 0$.

The stress-energy tensor responsible for two-lambda geometry connects
in a smooth way two vacuum states with non-zero cosmological constant:
de Sitter vacuum $T_{\mu\nu}={\kappa}^{-1} (\Lambda+\lambda)g_{\mu\nu}$
at the origin, and de Sitter vacuum $T_{\mu\nu}={\kappa}^{-1}\lambda
g_{\mu\nu}$ at infinity. This confirms the interpretation
of the stress-energy tensor with the algebraic structure 
(15) as corresponding to a variable 
effective cosmological term
$\Lambda_{\mu\nu}$ \cite{lambda}.

\section{De Sitter-Schwarzschild geometry}

\subsection{Horizons and objects}

De Sitter-Schwarzschild geometry (19) belongs to the class of  
solutions to the Einstein equations (8)-(10) with the source term 
$T_{\mu}^{\nu}$ of the algebraic structure (15). Formally it follows 
from imposing the condition $g_{00}=-g_{11}^{-1}$ on the metric \cite{me92}.
This condition applied to the system of equations (8)-(10),
follows in the condition $~T_t^t=T_r^r~$ which defines the class
of spherically symmetric geometries generated by a variable
cosmological term $~\Lambda_{\nu}^{\mu}$. De Sitter-Schwarzschild
geometry is the member of this class, distinguished by the boundary
conditions of de Sitter behaviour (12) at $~r=0~$ and Schwarzschild
behaviour (20) at infinity.

All the main features of this geometry follow from the algebraic
structure of a source term $\Lambda_{\mu\nu}$ and from the imposed
boundary conditions, independently of physical mechanism
responsible for $\Lambda_{\mu\nu}$, in the same way as all the main
features of de Sitter geometry are defined by the Einstein
cosmological term $\Lambda g_{\mu\nu}$, independently on physical 
mechanisms responsible for cosmological constant.

The fundamental difference from the Schwarzschild case is that
for any density profile satisfying (21)-(22), 
there are two horizons,
a black hole horizon $r_{+}$ and an internal Cauchy horizon $r_{-}$
 \cite{werner,valera,me92}.
A black hole horizon is formed by
outgoing radial photon geodesics - zero generators of a horizon. 
The Cauchy horizon consists of
ingoing radial photon geodesics which are not extendible to
the past. Therefore a global Cauchy surface does not exist
in this geometry.

Horizons are calculated as the positive roots of the equation $g_{00}(r)=0$.
 In the de Sitter-Schwarzschild geometry horizons exist 
for the masses in the region $M\leq M_{crit}$.
A critical value of the mass $M_{crit}$ marks the point at which 
the horizons come together. It gives the lower limit for
a black hole mass. 

Horizons are plotted in Fig.1
for the case of the density profile 
described by the Schwinger formula for
the vacuum polarization \cite{NF} in the spherically
symmetric gravitational field  which gives \cite{me92,me96}
$$
   \rho(r)={\kappa}^{-1}\Lambda \exp{\biggl(-\frac{\Lambda}{6GM}r^3\biggr)}=
        \rho_{vac} \exp{\biggl(-\frac{4\pi}{3}
               \frac{\rho_{vac}}{M}r^3\biggr)}\eqno(25)$$
The mass function in the metric (19) then takes the form
$$
    {\cal M}(r)=M\biggl(1-\exp{\biggl
       (-\frac{\Lambda}{6GM}r^3\biggr)}\biggr)\eqno(26)$$
The lower limit for a $\Lambda$BH mass is given by
$$
    M_{crit}\simeq{0.3M_{Pl}({\Lambda}_{Pl}/{\Lambda})^{1/2}}\eqno(27)$$
\begin{figure}
\vspace{-8.0mm}
\begin{center}
\epsfig{file=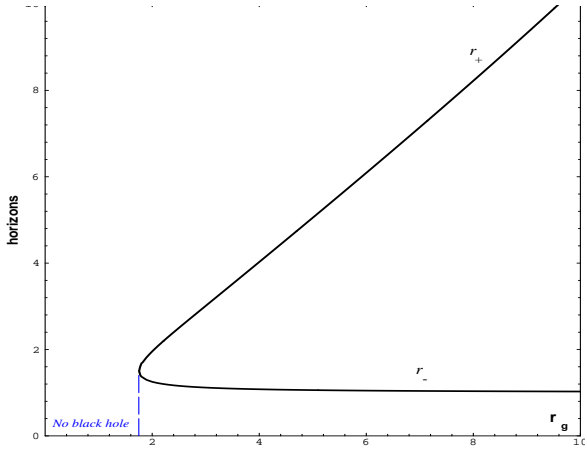,width=8.0cm,height=6.5cm,clip=}
\end{center}
\caption{
Horizons of de Sitter-Schwarzschild geometry 
plotted in the case of the density profile (25).}
\label{fig.1}
\end{figure}
Depending on the value of the mass $M$, there exist
three types of configurations 
in which a Schwarzschild singularity is replaced with $\Lambda$ core
\cite{me96,me99}:
1) a $\Lambda$ black hole  for $M>M_{crit}$;
2) an extreme $\Lambda$BH for $M=M_{crit}$; 
3) a "$\Lambda$ particle" ($\Lambda$P) - a particlelike structure without 
horizons made up of a self-gravitating spherically symmetric 
vacuum (15) - for $M < M_{crit}$.

De Sitter-Schwarzschild configurations are plotted in Fig.2  for the case
of the density profile (25).
\begin{figure}
\vspace{-8.0mm}
\begin{center}
\epsfig{file=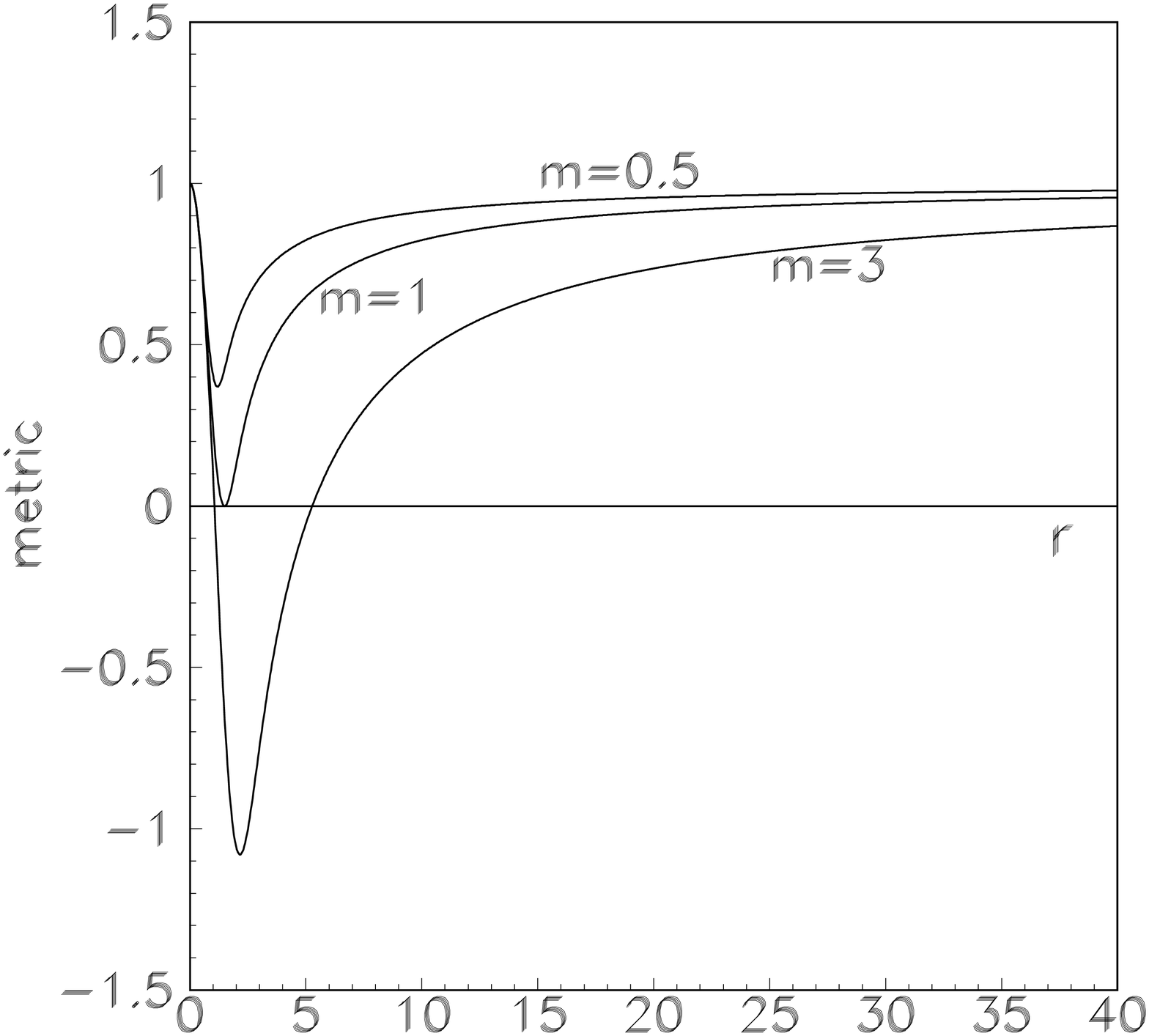,width=8.0cm,height=6.5cm}
\end{center}
\caption{
The metric coefficient $g_{00}(r)$ for de Sitter-Schwarzschild 
configurations in the case of the density profile (25). 
The parameter $m$ is the mass $M$ normalized to 
$M_{crit}\simeq{0.3M_{Pl}({\Lambda}_{Pl}/{\Lambda})^{1/2}}$.}
\label{fig.2}
\end{figure}
The question of the stability of a $\Lambda$BH and $\Lambda$P is currently
under investigation. Comparison 
of the ADM mass (21) with the proper mass 
$\mu$ 
which is the sum of the invariant masses of all particles \cite{MTW}
$$
   {\mu}=4\pi \int{{\rho(r)}(1-{2G{\cal M}(r)}/{r})^{-1/2}r^2dr}\eqno(28)$$
gives us a hint. In the spherically symmetric situations
the ADM mass represents the total energy, 
$M={\mu}+binding~ energy$ \cite{MTW}. 
In our case ${\mu}$ is bigger than $M$. This suggests 
that the configuration might be stable since energy is needed 
to break it up \cite{lambda}.

\subsection{Black and white holes}

The global structure of de Sitter-Schwarzschild spacetime in the case
$M>M_{crit}$ is shown in Fig.3 \cite{me96}.
It contains an infinite
sequence of black and white holes
whose singularities are replaced with future and past  
de Sitter regular cores ${\cal RC}$, 
and asymptotically flat universes ${\cal U}$.
Penrose-Carter diagram is plotted in coordinates of the photon radial
geodesics. The surfaces ${\cal J}^{-}$
and ${\cal J}^{+}$ are their past and future infinities. 
The surfaces $i^0$ are
spacelike infinities. The event horizons $r_{+}$ 
and the Cauchy horizons $r_{-}$ are formed by
the outgoing and ingoing radial photon geodesics $r_{\pm}$=const.
\begin{figure}
\vspace{-8.0mm}
\begin{center}
\epsfig{file=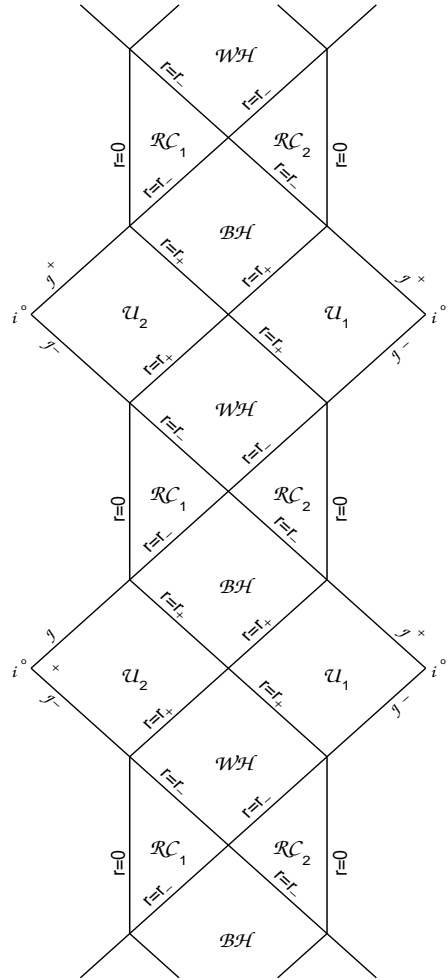,width=8.6cm,height=14.7cm} 
\end{center}
\caption{
Penrose-Carter diagram for $\Lambda$ black hole.}
\label{fig.3}
\end{figure}
It is evident from the Penrose-Carter diagram for $\Lambda$BH
that inside it there exists an inifinite number of
vacuum-dominated asymptotically flat universes 
in the future of $\Lambda$ white holes.
Geodesic structure of de Sitter-Schwarzschild space-time 
shows that the possibility of travelling into other universes through
a black hole interior, discussed in the literature for the case of
Reissner-Nordstrom and Kerr geometry (see, e.g., \cite{NF}), 
exists also in the case
of a $\Lambda$BH as the opportunity of safe travel - without a risk to
get into a singularity \cite{us00}. 
\vskip0.1in
{\bf $\Lambda$ white hole model for nonsimultaneous big bang-}
It is widely known that the interiors of black and white holes can be 
described locally as cosmological models (see, e.g., \cite{land}). 
In the case of a Schwarzschild white hole it starts from the
spacelike singularity $r=0$ (see Fig.4). 
\begin{figure}
\vspace{-8.0mm}
\begin{center}
\epsfig{file=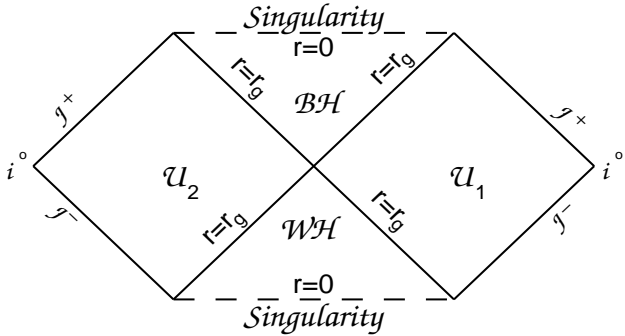,width=8.6cm,height=4.7cm} 
\end{center}
\caption{
Penrose-Carter diagram for the Schwarzschild black hole.}
\label{fig.4}
\end{figure}
Replacing a Schwarzschild singularity with the regular core $\cal{RC}$
trabsforms the spacelike singular surfaces $r=0$, both in the future
of a $\cal{BH}$ and in the past of a $\cal{WH}$, into the timelike
regular surfaces (see Fig.3). In a sense, this rehabilitates a white
hole whose existence in a singular version has been forbidden by
the cosmic censorship \cite{pen}.
 
Cosmological model related to a $\Lambda$WH corresponds to asymptotically
flat vacuum-dominated cosmology with the de Sitter origin, governed by
the time-dependent cosmological term $\Lambda_{\mu}^{\nu}$ (segment
$\cal{RC}$, $\cal{WH}$, $\cal{U}$ in the Fig.3). 
A $\Lambda$WH models thus initial stages of nonsingular cosmology
with an inflationary origin.

To investigate a $\Lambda$WH together with its past regular core
$\cal{RC}$ and future asymptotically flay universe $\cal{U}$,
we transform Schwarzschild coordinates ($t,r$)
into Finkelstein coordinates ($\tau,R$) related to raqial geodesics
of test particles at rest at infinity
$$
   c\tau=\pm\,ct\,\pm\,\int\sqrt{\frac{R_{g}(r)}{r}+f(R)}
      \frac{dr}{1-\frac{R_{g}(r)}{r}},\eqno(29) $$
$$
    R=ct+\int\sqrt{\frac{r}{R_{g}(r)}}\frac{\sqrt{1+f(R)}dr}
      {1-\frac{R_{g}(r)}{r}}\eqno(30)$$
Here 
$$
      R_g(r)=\frac{2G{\cal M}(r)}{c^2}\eqno(31)$$
and $f(R)$ is an arbitrary function satisfying the condition
$~ 1+f(R)>0$. The lower sign in (29) is for 
outgoing geodesics corresponding to the case of an expansion.

The metric (19) transforms into the Lemaitre type metric \cite{land}
$$
    ds^{2}=c^{2}d\tau^{2}-e^{\lambda(R,\tau)}
      dR^{2}-r^{2}(R,\tau)d\Omega^{2}\eqno(32)$$
with
$$
    e^{\lambda(R,\tau)}=\frac{R_{g}(r(R,\tau))}{r(R,\tau)}$$
Coordinates $R,\tau$ are the Lagrange (comoving) coordinates 
of a test particle, and $r$ 
is its Euler radial coordinate (luminosity distance). In the case ofoutgoing 
geodesics the $(R,\tau)$ coordinates with the lower sign in (29), 
map the segment ${\cal RC}, {\cal WH}, {\cal U}$, 
i.e. a $\Lambda$WH together with its past regular core ${\cal RC}$ 
and an external universe ${\cal U}$ in its absolute future.

For the metric (32) the Einstein equation reduce to \cite{land}
$$
   \kappa p_{r}=\frac{1}{r^{2}}\left(e^{-\lambda}r'^{2}-2r\ddot{r}-
      \dot{r}^{2}-1\right)\eqno(33)$$
$$
    \kappa p_{\perp}=\frac{e^{-\lambda}}{r}
      \left(r''-\frac{r'\lambda'}{2}\right)-
      \frac{\dot{r}\dot{\lambda}}{2r}-\frac{\ddot{\lambda}}{2}-
       \frac{\dot{\lambda}^{2}}{4}-\frac{\ddot{r}}{r}\eqno(34)$$
$$
   \kappa \rho=-\frac{e^{-\lambda}}{r^{2}}
      \left(2rr''+r'^{2}-rr'\lambda'\right)+\frac{1}{r^{2}}
       \left(r\dot{r}\dot{\lambda}+\dot{r}^{2}+1\right)\eqno(35)$$
$$
     \kappa T^{r}_{t}=\frac{e^{-\lambda}}{r}
       \left(2\dot{r}'-r'\dot{\lambda}\right)=0\eqno(36)$$
Here the dot denotes differentiation 
with respect to $\tau$ and ${}^{\prime}$ with respect to $R$. 
The component $T_{t}^{r}$ of the stress-energy tensor vanishes
in the comoving reference frame, and the Eq.(36) is integrated 
giving \cite{land}
$$
   e^{\lambda}=\frac{r'^{2}}{1+f(R)}\eqno(37)$$                               
Then we obtain from Eq.(33) the equation of motion
$$
    {\dot r}^2+2r{\ddot r}+\kappa p_r r^2=f(R)\eqno(38)$$
This cosmological model  
belongs to the Lemaitre class of spherically symmetric models with 
anisotropic fluid \cite{lem}. 
Dynamics of our model is governed by cosmological tensor
$\Lambda_{\mu\nu}$ which in this case is time-dependent.
For numerical integration of the equation of motion we adopt
the density profile (25).
Behavior of pressures $p_r^{\Lambda}=-\Lambda_r^r$ and
$p_{\perp}^{\Lambda}=-\Lambda_{\theta}^{\theta}=-\Lambda_{\phi}^{\phi}$ 
in this case is shown in Fig.5.
\begin{figure}
\vspace{-8.0mm}
\begin{center}
\epsfig{file=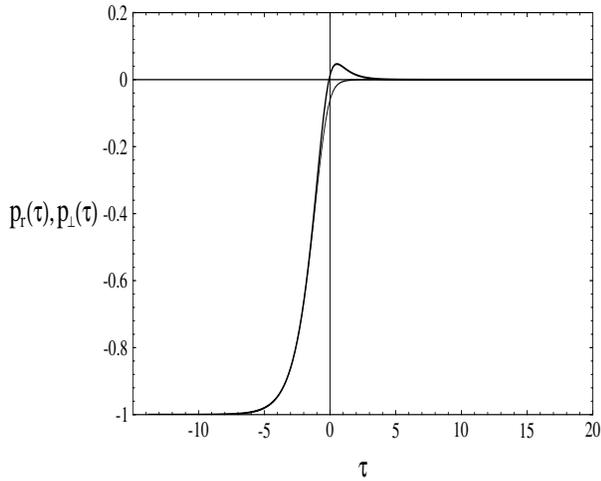,width=8.0cm,height=6.5cm}
\end{center}
\caption{
Radial and tangential pressures $p_r < p_{\perp}$.} 
\label{fig5}
\end{figure}
Near the surface $r=0$ the metric (32) transforms into
the FRW form with de Sitter scale factor for any $f(R)$.
The further evolution of geometry is calculated numerically.
Here we present numerical results for the case of $f(R)=0$ \cite{us99}
which correspond to the most popular today cosmological model with $\Omega=1$. 

The characteristic scale of de Sitter-Schwarzschild space-time
is $r_{*}=(r_0^2 r_g)^{1/3}$, and we normalize $r$ to this scale 
introducing dimensionless variable $\xi$ by $r=r_{*}\xi$. 
The equation of motion (38) for $f(R)=0$ reduces to 
$$
   \dot{\xi}^{2}+2\xi\ddot{\xi}-3\xi^{2}e^{-\xi^{3}}= 0\eqno(39)$$
It has the first integral
$$
   \dot{\xi}^{2}=\frac{A-e^{-\xi^{3}}}{\xi}\eqno(40)$$
and the second integral
$$
   \tau-\tau_{0}(R)=\int\limits_{\xi_{0}}^{\xi}\sqrt{\frac{x}
       {A-e^{-x^{3}}}}dx\eqno(41)$$
Here $\tau_{0}(R)$ is an arbitrary function (constant of integration 
parametrized by $R$) which is called the "bang-time function" \cite{silk}.
For example, in the case of the Tolman-Bondi model for a dust (an ideal
non-zero rest mass pressureless gas), the evolution is described by
$r(R,\tau)=(9GM(R)/2)^{1/3}(\tau-\tau_0(R))^{2/3}$, where $\tau_0(R)$
is an arbitrary function of $R$ representing the big bang singularity
surface for which $r(R,\tau)=0$ \cite{CS98}.

The bang starts from $\xi_{0}=0$ which is timelike regular 
surface (see Fig.3). 
Choosing $\xi_{0}=0$ we fix the constant $A=1$ and $\tau_0(R)=-R$. 
In coordinates $(R,\tau)$ bang starts from the surface $R+c\tau=-\infty$
shown in the Fig.6.
\begin{figure}
\vspace{-8.0mm}
\begin{center}
\epsfig{file=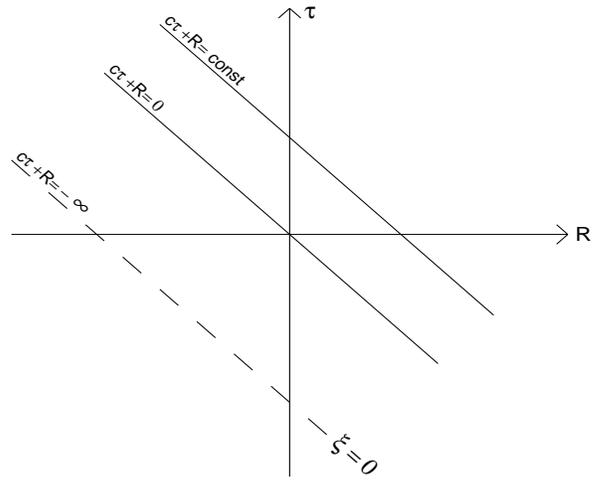,width=8.0cm,height=6.5cm}
\end{center}
\caption{
Surfaces $r=$const  plotted for dimensionless 
radius $\xi$. The surface $\xi=0$ is the big bang surface.}
\label{fig6}
\end{figure}
Different points of the bang surface $\xi=0$ start at different moments 
of synchronous time $\tau$.
 In the limit $\xi\rightarrow 0$, the metric takes the form
$$
    ds^{2}=c^{2}d\tau^{2}-r^{2}_{0}e^{\frac{2c\tau}{r_{0}}}
      \left(dq^{2}+q^{2}d\Omega^{2}\right)\eqno(42)$$ 
where the variable $q=e^{\frac{R}{r_{0}}}$ is introduced to transform the 
metric (32) into the FRW form. It describes, with the initial 
conditions $\xi_{0}(R+\tau\rightarrow -\infty)=0$, 
$\dot{\xi}_{0}(R+\tau\rightarrow -\infty)=0$, the nonsingular nonsimultaneous 
de Sitter bang.

In the case of a Schwarzschild WH, a singularity is spacelike
(see Fig.4),
so there exist the reference frame in which it is simultaneous.
In the case of a $\Lambda$ white hole, a surface $r=0$ is timelike
(Fig.3), and there does not exist any reference frame in which two events 
occuring on $r=0$ would be simultaneous.

The first lesson of the $\Lambda$WH model is that
nonsingular big bang must be nonsimultaneous \cite{us99}.

The further evolution of the function $\xi$,
velocity $\dot{\xi}$ and acceleration $\ddot{\xi}$
is shown in Figs 7-9, obtained by numerical integration of the equation of
motion (39) with the initial conditions $\xi_0=10^{-6}$,
$\dot{\xi}_0=10^{-6}$ \cite{us99}. 
\begin{figure}
\vspace{-8.0mm}
\begin{center}
\epsfig{file=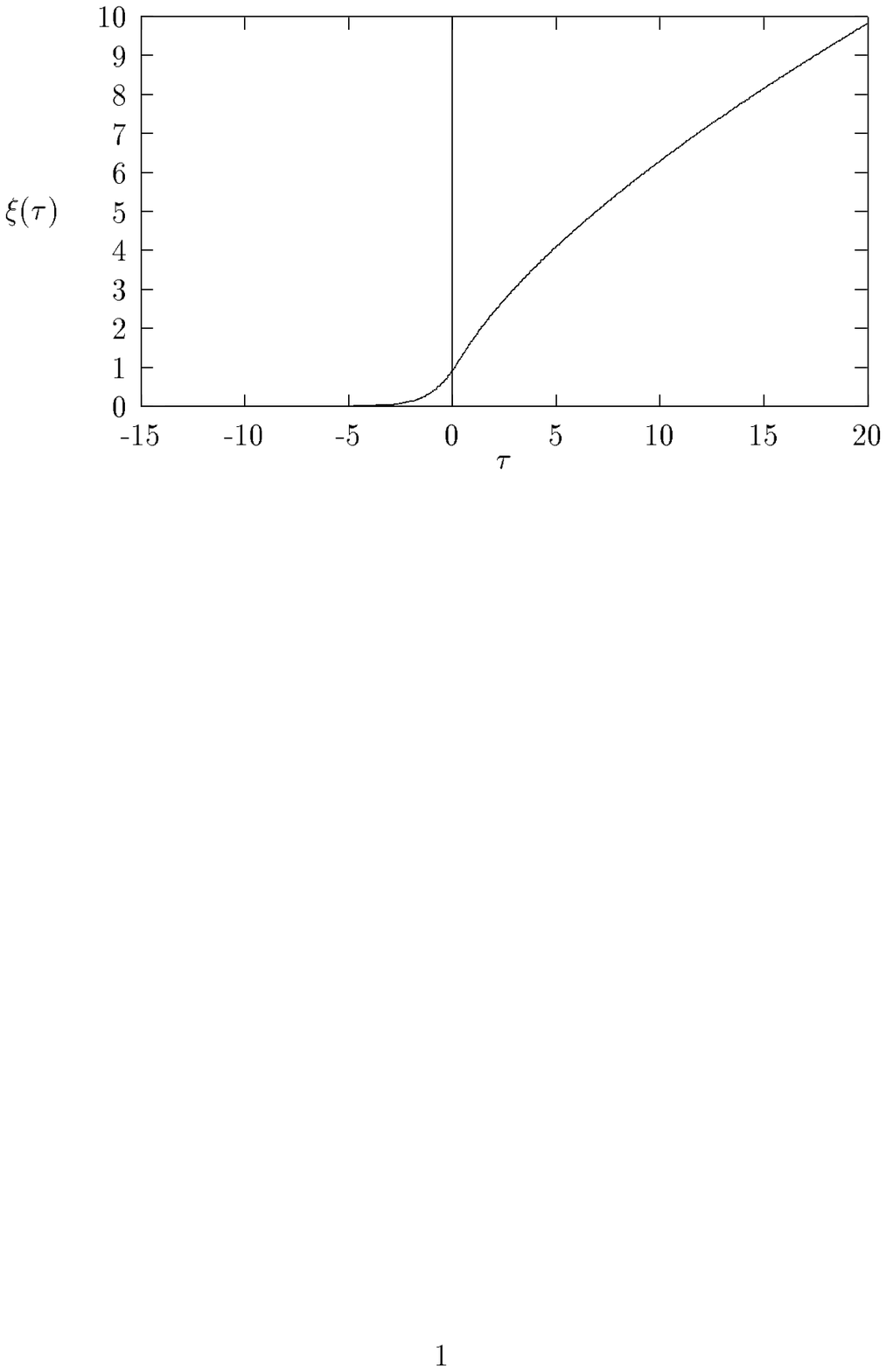,width=8.0cm,height=5.5cm}
\end{center}
\caption{
The function
$\xi(\tau - \tau_0)$ calculated from Eq.(39) 
with initial conditions ${\xi}_0= {\dot\xi}_0 = 10^{-6}$.}
\label{fig7}
\end{figure}
\begin{figure}
\vspace{-8.0mm}
\begin{center}
\epsfig{file=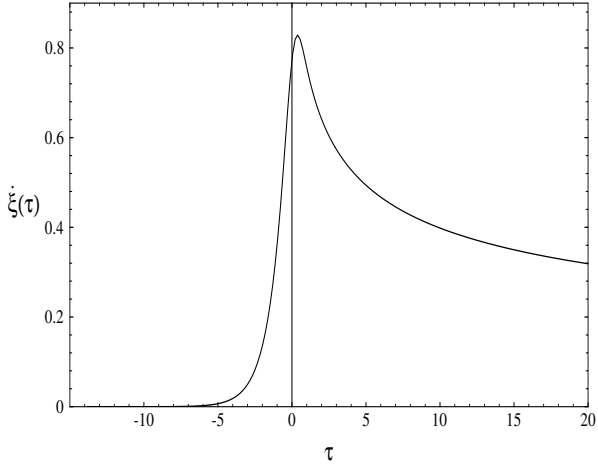,width=8.0cm,height=6.5cm}
\end{center}
\caption{
The plot of the velocity $\dot\xi(\tau - \tau_0)$ 
for initial conditions ${\xi}_0 = {\dot\xi}_0 = 10^{-6}$.}
\label{fig8}
\end{figure}
\begin{figure}
\vspace{-8.0mm}
\begin{center}
\epsfig{file=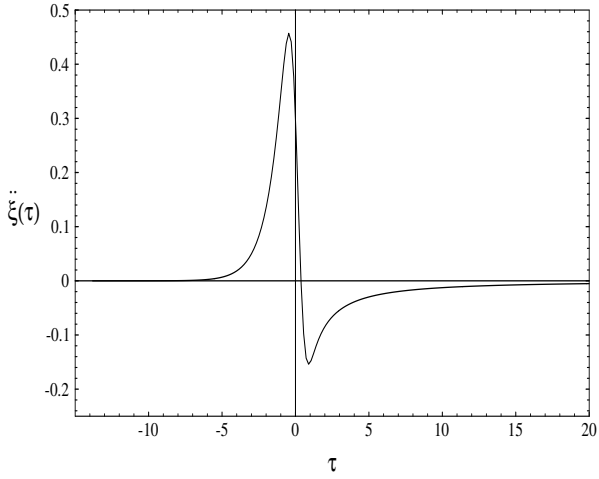,width=8.0cm,height=6.5cm}
\end{center}
\caption{
The acceleration of the "scale factor" $\xi(\tau-\tau_0)$.}
\label{fig9}
\end{figure}
Numerical integration of the equation (39) shows exponential growth 
of $\xi(\tau-\tau_0)$ at the begining, 
when $p_{\perp}\simeq p_{r}\simeq -\rho$, 
followed by anisotropic Kasner-type stage when different pressures lead to 
anisotropic expansion.

Qualitatively we can see this approximating the second integral (41)
in the region $1\ll{\xi}\ll (r_g/r_0)^{2/3}$ (far beyond a $\Lambda$WH
horizon) by
$$
     ds^{2}=c^{2}d\tau^{2}-\left(\frac{9r_{g}}{4}\right)^{
       \frac{2}{3}}\left(\tau+\widetilde{\tau}_{0}(R)\right)^{-\frac{2}{3}}
         \left(\frac{d\widetilde{\tau}_{0}(R)}{dR}\right)^{2}dR^{2}$$
     $$-\left(\frac{9r_{g}}{4}\right)^{\frac{2}{3}}\left(\tau
      +\widetilde{\tau}_{0}(R)\right)^{\frac{2}{3}}d\Omega^{2}\eqno(43)$$
where 
$\widetilde{\tau}_{0}(R)={R}+
\left(\frac{2}{3}\xi_{0}^{\frac{2}{3}}(R)-F(\xi_{0}(R))\right)$ 
and 
$F(\xi_{0}(R))=\int\limits_{0}^{\xi_{0}(R)}\sqrt{\frac
{x}{1-e^{-x^{3}}}}dx$.

This is anisotropic Kasner-type metric, with contraction in the radial 
direction, and expansion in the tangential direction.  Our model 
here differs from the Kasner vacuum solution by nonzero anizotropic pressures.

The second lesson of the $\Lambda$WH model is the existence
of the anisotropic Kasner-type stage after inflation.

This stage follows the nonsingular 
nonsimultaneous big bang from the regular surface $r=0$. 
It looks that this kind of behaviour is generic for cosmological models 
near the origin \cite{bkl} (for recent review see \cite{ber}). 
Our case differs from singular case also in that our solution displaying
Kasner-type stage is not vacuum in the sense of zero right-hand side
of the Einstein equations. However it is still vacuum-dominated
in the sense that the cosmological term $\Lambda_{\mu\nu}$ corresponds
to a spherically symmetric vacuum.

Since in our $\Omega=1$ model 3-curvature is zero, 
ADM mass ${\cal M}$ given by (18), coincides with 
the total proper mass $\mu$ given by (28) 
as the sum of the invariant masses of all particles 
with radial coordinate less then a certain value of $R$,
 frequently referred to as the Bondi mass \cite{bondi}.
At the beginning ${\cal M}=0$ and $\dot{\cal M}=0$, 
as it follows from the first integral (40) which gives \cite{us99}
$$
   \dot{\xi}^{2}=\frac{{\cal M}(\xi)}{\xi}\eqno(44)$$
The behavior of mass normalized to Schwarzschild mass $M$
is shown in the Fig.10.
\begin{figure}
\vspace{-8.0mm}
\begin{center}
\epsfig{file=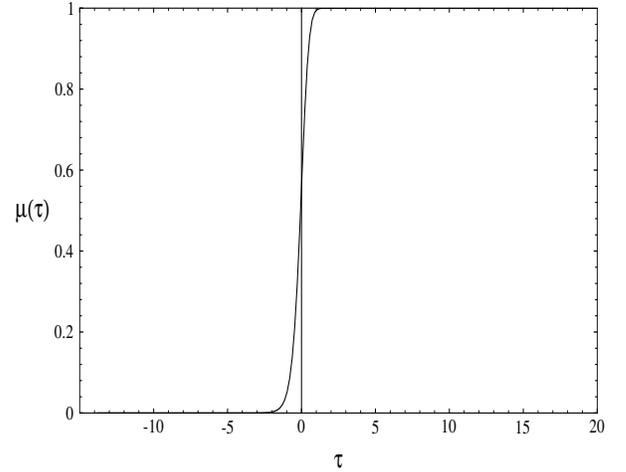,width=8.0cm,height=6.5cm}
\end{center}
\caption{
Plot of the mass function $\mu={\cal M}/M$.}
\label{fig10}
\end{figure}  
At the inflationary stage the mass increases as $\xi^{3}$. 
At the next anisotropic stage it is 
growing abruptly towards the value of the Schwarzschild mass $M$. 
Since the density is quickly falling at the some time 
starting from the initial value $\rho_{vac}={\kappa}^{-1} \Lambda$, the 
growth in a mass is connected with the fall of $\rho^{\Lambda}=\Lambda_t^t$,
i.e., with the decay of the initial vacuum energy (as it was noted
in the Ref \cite{us75}).

The third lesson of $\Lambda$WH model - quick growth of the mass 
during the Kasner-type anisotropic stage.

\vskip0.1in

{\bf Baby universes inside a $\Lambda$BH -}
In the case of direct matching of de Sitter to Schwarzschild 
metric \cite{fg,valera} the global structure 
of space-time corresponds to arising of a closed or semiclosed 
world \cite{valera} inside a BH (Fig.11).
\begin{figure}
\vspace{-8.0mm}
\begin{center}
\epsfig{file=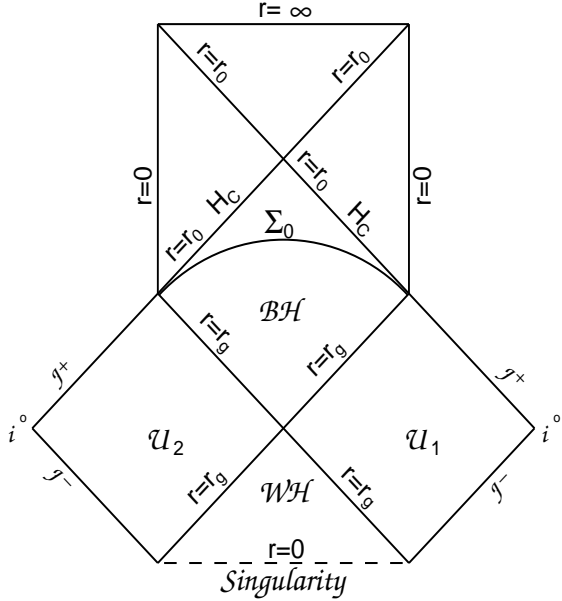,width=7.7cm,height=8.3cm}
\end{center}
\caption{
Penrose-Carter diagram for the case of the direct de Sitter-Schwarzschild
matching.} 
\label{fig11}
\end{figure}  
The conformal diagram shown in Fig.3 represents the global structure
of de Sitter-Schwarzschild spacetime in general case of a distributed density
profile. The situation near the surface $r=0$ is similar to the 
case considered by Farhi and Guth \cite{fg,fgg}.
The region near $r=0$, which  is the part of the regular core 
${\cal RC}$, differs from that considered in Ref.\cite{fgg}
by an $r$-dependent density profile. Our density profile (25)
is almost constant near $r\rightarrow 0$ and then quickly
falls down to zero. In Fig.12 it is plotted for the case
of a stellar mass black hole with $M=3M_{\odot}$. 
\begin{figure}
\vspace{-8.0mm}
\begin{center}
\epsfig{file=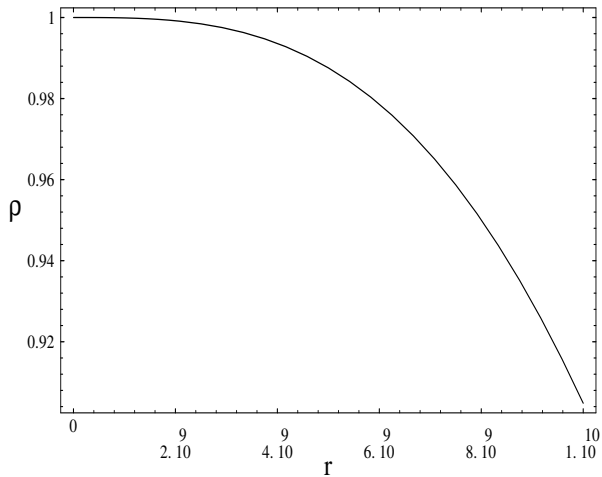,width=8.0cm,height=6.5cm}
\end{center}
\caption{
Density profile (25) for the case of 
of three solar masses nonsingular black hole.}
\label{fig12}
\end{figure}  
We may think of
the region near $r=0$ as of a small false vacuum bubble
which can be a seed for a quantum birth of a new universe.
In this context quantum creation of 
a baby universe inside a $\Lambda$BH would change the global structure 
of spacetime as it is shown in Fig.13, which corresponds to arising 
of a closed or semiclosed world in one of $\Lambda$WH structures 
in the future of a $\Lambda$BH in the original universe.
\begin{figure}
\vspace{-8.0mm}
\begin{center}
\epsfig{file=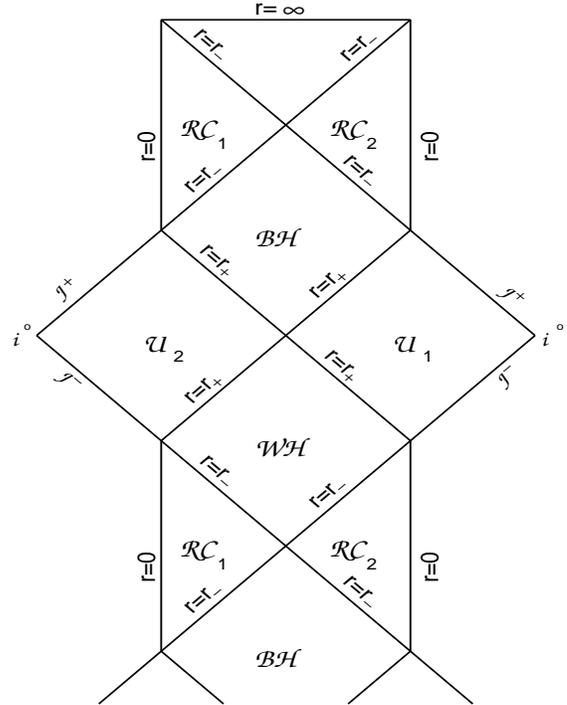,width=7.7cm,height=9.7cm}
\end{center}
\caption{
The global structure of space-time for the case of creation of 
closed or semiclosed universe inside a $\Lambda$ black hole.}
\label{fig13}
\end{figure} 
On the other hand, in the case of a $\Lambda$BH with a distributed
density profile, whose global structure (Fig.3) is essentially
different from that of direct matching case (Fig.11), 
there exist  possibilities other than considered by Farhi and Guth.

It has been well known since the very beginning of studying white holes
that they are unstable. Instabilities of  Schwarzschild white holes
are related to physical processes (particle creation) near a singularity
(see \cite{NF} and references therein).
In the case of a $\Lambda$WH its quantum instability  
is related to instability of de Sitter vacuum 
near the surface $r=0$. 
Therefore we can consider arising of 
a baby universe in a $\Lambda$BH as the result of quantum 
instability of de Sitter vacuum.

Instability of de Sitter vacuum is well studied, bith with
respect to particle creation (see \cite{bir,serg}), and with
respect to quantum birth of 
a universe \cite{us75,gott,lin,as,vil,vil2,linb,olive}
The possibility of birth of a universe from de Sitter vacuum 
has been widely discussed in the literature. 

The possibility of
multiple birth of causally disconnected universes 
from de Sitter background has been noticed in our 1975
paper \cite{us75}. In 1982 such a possibility
has been investigated by J. Richard Gott III who considered creation 
of a universe as a quantum barrier penetration leading to
an open FRW cosmology \cite{gott}. 
Linde \cite{lin} and Albrecht and Steinhardt \cite{as}
have suggested detailed mechanisms for forming bubbles, and now
these single-bubble models are spoken of as the "new inflationary
scenario" (for review see \cite{linb}).

The case of arising of an open or flat universe from de Sitter vacuum 
is illustrated by Fig.14 from J. Richard 
Gott III paper \cite{gott}. The events $E$ and $E^{\prime}$ are birth
of causally disconnected universes (open in Gott III case) 
from de Sitter vacuum. The curved
lines are world lines of comoving observers. At the spacelike surface
$AB$ the phase transition occurs from the inflationary 
$p=-\rho$ stage to the radiation dominated $p=\rho/3$
stage \cite{us75,gott}. 

In the case of a $\Lambda$BH the region $ECB$
in Fig.14 corresponds to the region ${\cal{RC}}_1$ in Fig.3, and
the region $BFD$ corresponds to the part of region ${\cal{RC}}_2$.
Those two regions in de Sitter-Schwarzschild space-time are
entirely disjoint from each other for the same reasons as the
regions ${\cal{U}}_1$ and ${\cal{U}}_2$ (they can be connected
only by spacelike curves). Birth of baby universes
inside a $\Lambda$BH looks very similar to the picture shown
in Fig.14.
\begin{figure}
\vspace{-8.0mm}
\begin{center}
\epsfig{file=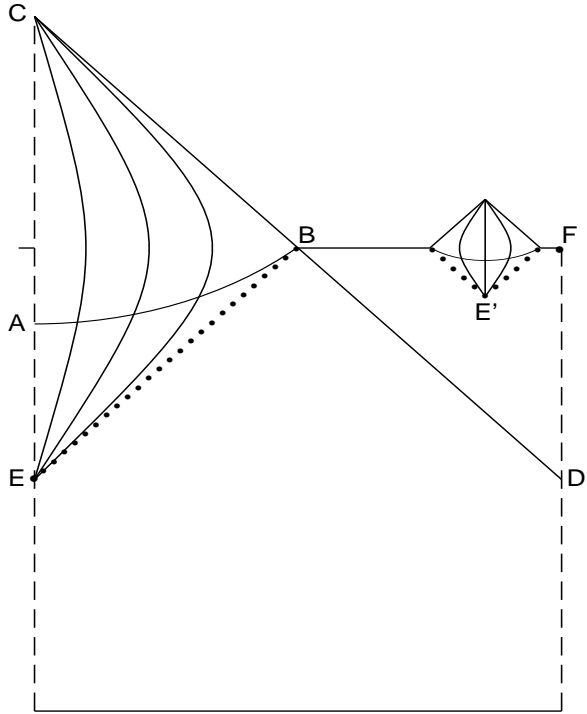,width=8.0cm,height=9.8cm}
\end{center}
\caption{
Penrose-Carter diagram \protect\cite{gott}, which corresponds to the case 
of birth of  baby universes inside a $\Lambda$BH \protect\cite{us99}.}
\label{fig14}
\end{figure} 
In any case a nucleating bubble is spherical and can be
described by a minisuperspace model with a single
degree of freedom \cite{vil2}, in our case $a=(r_0^2r_g)^{1/3}\xi$.

The Friedmann equation in the conformal time ($cdt=ad\eta$) reads 
$$
    \biggl(\frac{da}{d\eta}\biggr)^2
      =\frac{8\pi G\rho a^4}{3c^2}-ka^2,\eqno(45)$$           
where $k=0,\pm 1$. Standard procedure of quantization \cite{vil,vil2}
results in the Wheeler-DeWitt
equation in  the minisuperspace for the wave function of universe \cite{vil}
$$
    \frac{d^2 \psi}{da^2}-V(a)\psi=0,\eqno(46)$$
where for the case of $k=1$
$$
    V(a)=\frac{1}{l_{Pl}^4}\biggl(a^2 - \frac{a^4}{r_0^2}\biggr)\eqno(47)$$
where $r_0$ is given by (13).
With this equation we can estimate the probability of tunneling event
describing the quantum growth of an initial
bubble on its way to the classically permitted region $a\geq r_0$,
which corresponds to the case of a closed universe inside a black hole
as in models of Refs\cite{fg,valera}.

Eq. (46) reduces to the Schr$\ddot{o}$dinger equation 
$$
   \frac{{\hbar}^2}{2m_{Pl}}\frac{d^2 \psi}{da^2}-[U(a)-E]\psi=0\eqno(48)$$
with $E=0$ and
$$
   U(a)=\frac{m_{Pl}c^2}{2{l_{Pl}}^2}\biggl(a^2
     -\frac{a^4}{r_0^2}\biggr)\eqno(49)$$
This potential is plotted in Fig.15. It has two zeros, at $a=0$ and $a=r_0$, 
and two extrema: the minimum at $a=0$ and the maximum at $a=r_0\sqrt{2}$.
\begin{figure}
\vspace{-8.0mm}
\begin{center}
\epsfig{file=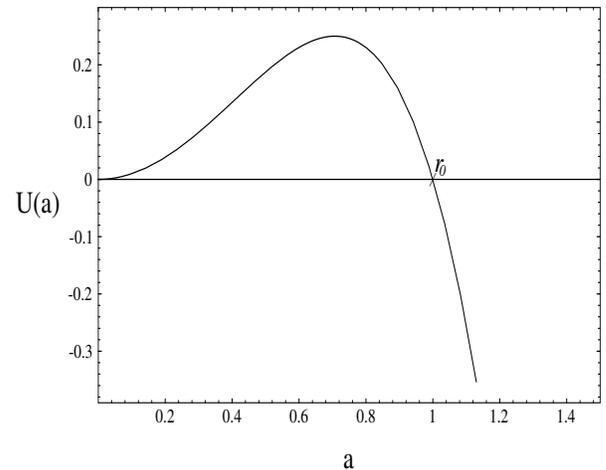,width=8.0cm,height=6.5cm}
\end{center}
\caption{
Plot of the potential Eq.(49).}
\label{fig15}
\end{figure}  
The WKB coefficient for penetration through the potential barrier reads
\cite{ll}
$$
   D=\exp{\bigg(-\frac{2}{\hbar}|\int_{a_1}^{a_2}
        {\sqrt{2m_{Pl}[E-U(a)]}da}|\biggr)}\eqno(50)$$
which gives
$$
     D=\exp{\biggl[-\frac{2}{3}\biggl(\frac{r_0}{l_{Pl}}
        \biggr)^2\biggr]}\eqno(51)$$
in agreement with the results of Ref.\cite{fgg}.

The probability of a single tunneling event
is very small. 
For the Grand Unification scale $\sim{10^{15}}$ GeV this
probability is of order of $\exp(\frac{2}{3}10^{16}$ (which agrees
with the Farhi, Guth and Guven result \cite{fgg}).
However in our case there exists an infinite set of $\Lambda$WH 
structures inside the $\Lambda$BH. This strongly magnifies the probability
of birth of a baby universe, sooner or later, in one of them. 

To estimate probability of quantum birth of an open or flat universe
we have to take into account possibility of the equation of state
other than $p=-\rho$ in the initial small seed bubble.
We can consider the instability 
of a $\Lambda$WH as evolved from a quantum fluctuation near $r=0$. 
It is possible to find within this fluctuation, 
some admixture of quintessence (spatially inhomogeneous component 
of matter content with negative pressure \cite{CDS}) 
with the equation of state $p=-\rho/3$.
In this case it is possible to find nonzero
probability of tunneling for any value of $k$ \cite{F}.

In the Friedmann equation (45)
 the density evolves with the scale factor $a$ as
$$
   \rho=\rho_{vac}\biggl(\frac{a}{r_0}
        \biggr)^{-3(1+\alpha)},\eqno(52)$$
where
$\alpha$ is a factor in the equation of state 
$p=\alpha\rho$. For the de Sitter vacuum $\alpha=-1$, 
and $\alpha=-1/3$ for the equation of state $p=-\rho/3$.
When both those  components are
present in the initial fluctuation, the  density can be 
written in the form \cite{F}
$$
   \rho=\rho_{vac}\biggl(B_0
          +B_2\frac{r_0^2}{a^2}\biggr),\eqno(53)$$
where $B_0$ and $B_2$ refer to the corresponding contributions.

Then the Friedmann equation (45) takes the form
$$
   \biggl(\frac{da}{d\eta}\biggr)^2=(B_2-k)a^2+\frac{B_0a^4}{r_0^2}\eqno(54)$$
which transforms to the  Schr\"odinger equation (48)
with zero energy and with the potential 
$$
          U(a)=\frac{m_{pl}c^2}{2l_{pl}^2}\biggl[(k-B_2)a^2
            -\frac{B_0a^4}{r_0^2}\biggr]\eqno(55)$$
This potential has two zeros at $a_1=0$ and $a_2=\sqrt{(k-B_2)/{B_0}}r_0$ 
and two extrema: the minimum
at $a=0$ and the maximum at $a=r_0\sqrt{(k-B_2)/(2B_0)}$.

The WKB coefficient for penetration through the potential barrier reads
now \cite{us99}
$$
       D=exp\left(-\frac{2}{l_{Pl}}|\int\limits_{a_{1}}^{a_{2}}
        \sqrt{(k-B_2)a^2-\frac{B_0 a^4}{r_0^2}}da|\right)\eqno(56)$$
The presence of quintessence with the equation
of state $p=-\rho/3$ in the initial fluctuation  provides
a possibility of quantum birth for the case of both 
flat $(k=0$) and open ($k=-1$)
universe inside a black hole.
The probability of the birth of universe 
for this case is given by \cite{us99}
$$
          D=\exp{\left(-\frac{2}{3}\left(\frac{r_0}{l_{pl}}\right)^2
          \frac{\sqrt{(k-B_2)^3}}{B_0}\right)}\eqno(57)$$
For $r_0\sim 10^{-25}$ cm, $D=\exp{(-\frac{1}{3}\cdot 10^{16})}$ 
for $k=0$, $B_0=2$, $B_2=-1$.
This is very close to the value calculated above for the case
of $k=1$ and $B_2=0$, and to that obtained by Farhi, Guth and Guven
\cite{fgg}. And, as was pointed above, this small probability of
a single tunneling event is magnified in the case of a $\Lambda$BH
by the infinite number of appropriate $\Lambda$WH structures
inside of each particular $\Lambda$BH.

Let us emphasize, that an obstacle related to the initial singularity, 
does not arise in general case of distributed profile, since both future 
and past singularities are replaced by the regular surfaces $r=0$.
On the other hand, in the context of creation of a universe 
in the laboratory, the possibility of influence on the new universe 
is restricted by the presence of the Cauchy horizon in 
the de Sitter-Schwarzschild geometry.

\subsection{Thermodynamics of $\Lambda$BH}

De Sitter-Schwarzschild black hole emits Hawking radiation from both horizons, 
with the Gibbons-Hawking temperature \cite{gh}
$$
          T=\frac{\hbar\Upsilon}{2\pi kc},\eqno(58)$$
where $k$ is the Boltzmann constant.
The surface gravity $\Upsilon$ satisfies the equation 
$$
        K_{a;b}K^b=\Upsilon K_a\eqno(59)$$
on the horizons, where $K_a$ is the Killing vector normalized to have unit
magnitude at the origin. It is uniquely defined by the conditions
that it should be null on both horizons and have unit magnitude 
at $r\rightarrow\infty$. For de Sitter-Schwarzschild black hole 
the surface gravity is given by \cite{me96}
$$
      \Upsilon=\frac{c^2}{2}\biggl[\frac{R_g(r_h)}{{r_h}^2}-
          \frac{R_g^{\prime}(r_h)}{r_h}\biggr]\eqno(60)$$
with $r_h$ for both $r_{+}$ and $r_{-}$.
It has biffurcation point at $M=M_{crit}$ (extreme black hole)
determined by the condition $g_{00}(r_{\pm})
=g^{\prime}_{00}(r_{\pm})=0$ and given by (27).

For ${\cal M}(r)$ defined by (18) the Hawking temperature is \cite{me96}
$$
     T_h=\frac{\hbar c}{4\pi kr_0}\biggl[\frac{r_0}{r_h} -\frac{3r_h}{r_0}
           \biggl(1-\frac{r_h}{r_g}\biggr)\biggr].\eqno(61)$$
Surface gravity characterizes the force that must be 
exerted at infinity to hold a unit test mass in place at the horizon. It has 
opposite signs for $r_{+}$ and $r_{-}$ due to repulsive character of gravity 
near $r_{-}$. Therefore the temperature on the internal horizon is negative.

The temperature-mass diagram is shown in the Fig.16.
\begin{figure}
\vspace{-8.0mm}
\begin{center}
\epsfig{file=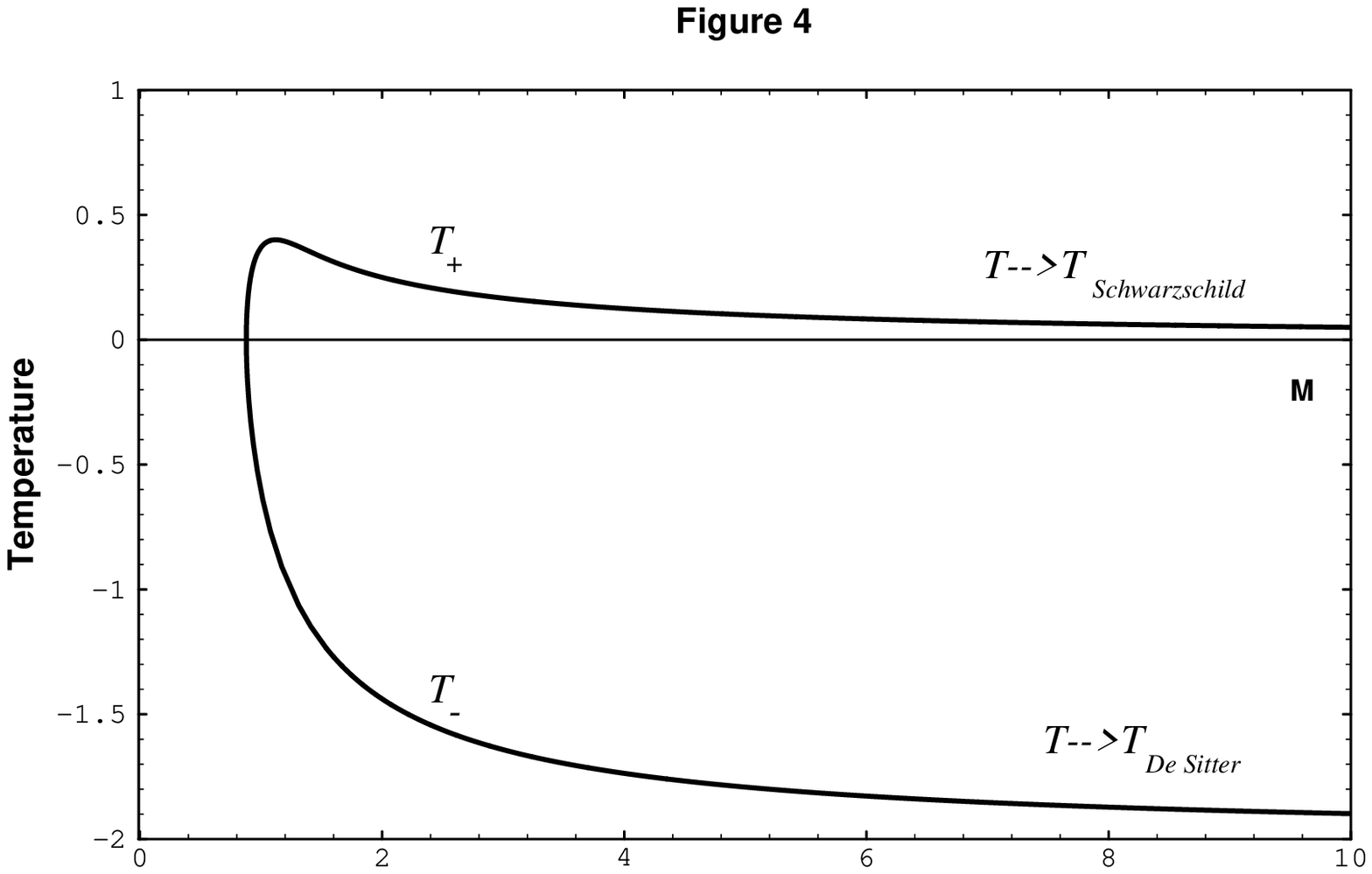,width=8.0cm,height=6.5cm,clip=}
\end{center}
\caption{The temperature-mass diagram for $\Lambda$BH
in units $\hbar=c=G=8\pi k=1$, for the external and internal
horizons. The specific heat on the external horizon
is positive for $M_{crit}\leq M \leq M_{tr}$, 
where $M_{crit}\approx 0.3 M_{Pl} (\Lambda_{Pl}/\Lambda)^{1/2}$ and
$M_{tr} \approx 0.38 M_{Pl} (\Lambda_{Pl}/\Lambda)^{1/2}$. A second-order
phase transition occurs at $M=M_{tr}$.}
\label{fig16}
\end{figure}  
For $r_g\gg r_0$ the temperature tends to the Schwarzschild value 
$$
         T_{Schw}=\frac{\hbar c^3}{8\pi k GM}\eqno(62)$$  
on the external horizon and to the de Sitter value 
$$
         T_{deS}=-\frac{\hbar c}{2\pi k r_0}\eqno(63)$$  
on the inner horizon. 
Temperature has the maximum at
$$
    M_{tr}\simeq{0.38 M_{Pl}({\Lambda}_{Pl}/{\Lambda})^{1/2}},\eqno(64)$$
and drops to zero when $r_h$ approaches $r_{+}=r_{-}$
as $M$ approaches $M_{crit}$. As $M$ decreases within range of masses 
$M_{tr}\geq M\geq M_{crit}$, the temperature $T_{+}$ also decreases. 
The specific heat is positive in this mass range.
In the range of masses $M>M_{tr}$ temperature grows 
as black hole loses its mass in the course
of Hawking evaporation, approximately according to (62).
The specific heat is negative in this mass range. 
 It changes sign as the mass 
decreases beyond $M_{tr}$, hence a second-order phase 
transition occurs at this point.
Temperature of a phase transition is given by
$$
      T_{tr}\simeq{0.2 T_{Pl}({\Lambda}_{Pl}/{\Lambda})^{1/2}}\eqno(65)$$
For the case of ${\Lambda}\sim{{\Lambda}_{GUT}}$ and $M_{GUT}\sim{10^{15}GeV}$,
$$
    T_{tr}\sim{0.2\times{10^{11}GeV}}$$ 
        $$M_{tr}\sim{0.4\times{10^{11}GeV}};~ 
          M_{crit}\sim{0.3\times{10^{11}GeV}}$$
Existence of the lower limit for a $\Lambda$BH mass 
follows in fact from existence of two horizons and
does not depend ob particular form for the density profile.
Temperature drops to zero when horizons merge at a certain value of mass 
$M_{crit}$, and vanishes as $T\sim {M^{-1}}$ for $M\gg M_{crit}$. 
It is nonzero and positive for $M_{crit}<M<\infty$, since surface gravity 
is nonzero and positive on the external horizon, in accordance with the 
laws of black hole thermodynamics. 
Therefore a curve representing temperature-mass dependence on the external 
horizon must have a maximum at a certain value of mass $M_{tr}>M_{crit}$, 
which means a second-order phase transition at $M=M_{tr}$
\cite{me96,haifa}.

The vacuum energy outside a $\Lambda$BH horizon is given by
$$
         E_{vac}=\int_{r_{+}}^{\infty}{\rho(r)r^2dr}=
          M\exp{\biggl(-\frac{\Lambda}{6GM}r_{+}^3\biggr)}\eqno(66)$$
One can say that $\Lambda$BH has $\Lambda$ hair 
vanishing only asymptotically as $(r_g/r_0)\rightarrow\infty$. 

In the course of Hawking evaporation, a $\Lambda$BH loses its mass
and the configuration evolves towards a $\Lambda$P \cite{me96}. 

\subsection{Characteristic surfaces of de Sitter-Schwarzschild geometry}

De Sitter-Schwarzscild spacetime has two characteristic surfaces
at the characteristic scale $r\sim{(r_0^2r_g)^{1/3}}$. 

The surface of zero gravity is the characteristic surface 
at which the strong energy condition of singularities theorems \cite{haw} 
$(T_{\mu\nu}-g_{\mu\nu}T/2)u^{\mu}u^{\nu}\geq 0$
(where $u^{\nu}$ is any timelike vector) is violated, which means that 
gravitational acceleration changes sign. The radius of this surface satisfies 
the condition $\varepsilon +\sum{p_k}=0$, where $p_k=-T_k^k$, and is given by 
$$
         r=r_c={(2r_0^2r_g/3)}^{1/3}\eqno(67)$$

The globally regular configuration with the de Sitter core instead of a
singularity arises as a result of balance between attractive gravitational
force outside and repulsive gravitational force inside of the region 
bounded by the surface of zero gravity (67).      

The surface of zero scalar
curvature $R=0$ is  
          $$r=r_s={(4r_0^2r_g/3)}^{1/3}\eqno(68)$$
Beyond this surface the scalar curvature changes sign. 

Choosing $r_s$ as the characteristic size of a regular core,
we can estimate its gravitational (ADM) mass (18) by
$$
     m_{core}=4\pi \int_0^{r_s}{\rho r^2 dr}\simeq{0.74 M}\eqno(69)$$
Let us note that this mass is confined within the region
$$
         r_s\simeq10^{-20}cm ({\Lambda}_{Pl}/{\Lambda})^{1/3}
         (M/3M_{\odot})^{1/3}\eqno(70)$$
For a black hole of several solar masses it means that 
 $\sim{{10^{34}g}}$ is contained in the core
of a size $r_s < 10^{-15}cm$.

Four characteristic surfaces $r_{+}$, $r_{-}$, $r_c$ and $r_s$ 
are plotted in Fig.17. 
\begin{figure}
\vspace{-8.0mm}
\begin{center}
\epsfig{file=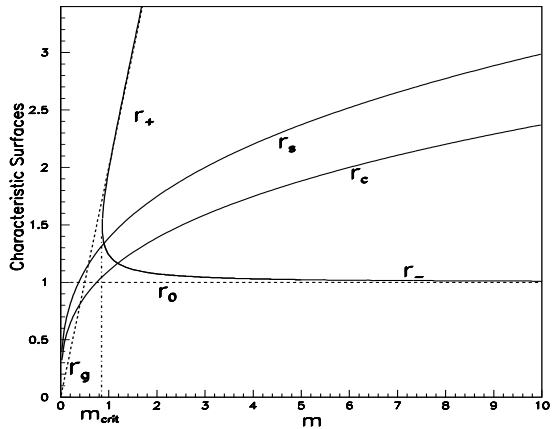,width=8.0cm,height=6.5cm}
\end{center}
\caption{Surface of zero scalar curvature $r=r_s$ and of zero gravity
$r=r_c$ (strong energy condition of singularity theorems is violated)
are shown together with horizons.}
\label{fig17}
\end{figure}  
Characteristic surfaces of de Sitter-Schwarzschild geometry
exhibit nontrivial behaviour: as mass parameter  
decreases they change their places before horizons merge, 
suggesting nontrivial dynamics of evaporation \cite{haifa}.
For nonextreme black hole characteristic surfaces $r_c$ and $r_s$ are
located between horizons $r_{+}, r_{-}$, in such a sequence that
$$
        r_{-}<r_c<r_s<r_{+}\eqno(71)$$
At the value of mass  $M_c\simeq{0.41 M_{Pl}
({\Lambda}_{Pl}/{\Lambda})^{1/2}}$, the surface of zero gravity
$r=r_c$ merges with the internal horizon $r_{-}$, and then
for masses $M<M_c$, the sequence becomes
$$
    r_c<r_{-}<r_s<r_{+}\eqno(72)$$
At the value of mass  $M_s\simeq{0.31 M_{Pl}
({\Lambda}_{Pl}/{\Lambda})^{1/2}}$, 
the surface of zero scalar curvature $r=r_s$ merges with the
internal horizon, and for masses $M<M_s$
$$ 
        r_c<r_s<r_{-}<r_{+},\eqno(73)$$
For the extreme black hole
$$
         r_c<r_s< r_{-}=r_{+},\eqno(74)$$
and for recovered particlelike structure without horizons
$$
        r_c<r_s\eqno(75)$$
In accordance with results of Ref.\cite{maeda}, the de Sitter-like core 
can approximate the non-Abelian structure with mass $M=\sqrt 2 gv$, 
where $v$ is the vacuum expectation value of the relevant
non-Abelian (Higgs) field with selfinteraction $\lambda=g^2$. 
We can estimate a coupling constant $g$, setting the Compton wavelength 
equal to the characteristic size $r_s$ of a particlelike structure.  
It results in the formula for the mass
$M=\sqrt 2 gv=M_{Pl}(2\pi\Lambda/{\Lambda_{Pl}})^{1/4}$. 
Taking into account that
$(\Lambda/\Lambda_{Pl})=(v/M_{Pl})^4$, we obtain $g=(\pi/4)^{1/4}$.
Corresponding fine-structure constant $\alpha=g^2/{4\pi}$ is
$\alpha=1/(8\sqrt{\pi})\approx{1/14}$ \cite{me96}.

\subsection{Vacuum cores of fundamental particles}

De Sitter-Schwarzschild particlelike structure cannot be applied
straightforwardly to approximate a structure of a fundamental particle
like an electron which is much more complicated. However,
we can apply this model to estimate lower limits for sizes of FP, 
in the frame of our assumption that a mass of a FP is related to its 
gravitationally induced core with de Sitter vacuum $\rho_{vac}$ at $r=0$. 
This allows us to estimate a lower limit on a size of a FP by the size of its 
vacuum core $r_c$ defined by the de Sitter-Schwarzschild geometry.

 We do this in two cases \cite{zur}. First is the
case when a lepton gets its mass via the Higgs mechanism at the
electroweak scale. Second is the case of maximum possible $\rho_{vac}$
permitted by causality arguments for a given mass, independently on
mechanism of mass generation. 

In the context of spontaneous symmetry breaking the false vacuum
density $\rho_{vac}$ is related to the vacuum expectation value
$v$ of a scalar field which gives particle a mass $m=gv$,
where $g$ is the relevant coupling to the scalar. For a scalar particle
$g=\sqrt{2\lambda}$ where now $\lambda$ is its self-coupling. It is neutral 
and spinless, and we can approximate it by de Sitter-Schwarzschild particlelike
structure - in accordance with EYMH results \cite{maeda}. We identify then
the vacuum density $\rho_{vac}$ of the self-gravitating object with
the self-interaction of the Higgs scalar in the standard $\phi^4$ theory
$$
     \rho_{vac}=\frac{\lambda v^4}{4}\eqno(76)$$
Then the size of a vacuum inner core for a lepton with the mass $m_l$
is given by
$$
   r_c=\biggl(\frac{2m_l}{\pi\lambda v^4}\biggr)^{1/3}\eqno(77)$$
We assume also that the gravitational size of a particle $r_s$
confining most of its mass, is restricted by its Compton wavelength.
This is the natural assumption for two reasons: first, for  a quantum
object Compton wavelength constraints the region of its localization.
Second, it follows from all experimental data concerning limits on sizes
of fundamental particles measured by characteristic sizes of interaction
regions. For example, for the process 
$e^{+}e^{-}\rightarrow \gamma\gamma(\gamma)$
at energies around 91 GeV and 130-183 GeV using the data collected with
the L3 detector from 1991 to 1997 (for review see \cite{zur}) all
the sizes of interaction regions are found to be smaller than the Compton
wavelength of FP \cite{zur}.

Appplying this constraint for a Higgs scalar we
get for its self-coupling $\lambda\leq\frac{\pi}{16}$.
Vacuum expectation value for the electroweak scale
is $v=246$ GeV, and an upper limit on a mass of a scalar 
is given by \cite{zur}
$$
     m_{scalar}\leq 154~ GeV\eqno(78)$$ 
The lower limits for sizes of leptons vacuum cores
for electron, muon and tau are estimated to \cite{zur} 
$$
   r_c^{(e)}>1.5\times{10^{-18}cm},~~~r_c^{(\mu)}>0.9\times{10^{-17}cm},$$
   $$~~~r_c^{(\tau)}>2.3\times{10^{-17}cm}\eqno(79)$$
These numbers are close to experimental constraints. For example,
for electron the characteristic size $R_e$ is restricted by
$e^{+}e^{-}\rightarrow \gamma\gamma(\gamma)$ reaction to 
$R_e<1.5\times{10^{-17}cm}$, and the direct contact term measurements
for the electroweak interaction constrain the characteristic size for the
leptons to $R_l<2.8\times{10^{-18}cm}$ \cite{zur}.

To estimate the most stringent limit on $\rho_{vac}$
we take into account that quantum region of localization given
by Compton wavelength must fit within a causally connected region
confined by the de Sitter horizon $r_0$. This requirement gives the limiting
scale for a vacuum density $\rho_{vac}$ related to a given mass $m$
$$
      \rho_{vac}\leq{\frac{3}{8\pi}}\biggl
           ({\frac{m}{m_{Pl}}}\biggr)^2\rho_{Pl}\eqno(80)$$
This condition connects a mass $m$ with the scale for a vacuum density
$\rho_{vac}$ at which this mass could be generated in principle,
whichever would be a mechanism for its generation.
In the case if FP have inner vacuum mass cores generated at the
scale of Eq.(80) (for the electron this scale is of order of 
$4\times{10^7}$ GeV), we get the most stringent, model-independent
lower limit for a size of a lepton vacuum core
$$
  r_c>\biggl({\frac{4}{3}}\biggr)^{1/3}\biggl({\frac{m_{Pl}}{m_l}}
      \biggr)^{1/3}l_{Pl}\eqno(81)$$
It gives for electron, muon and tau the constraints \cite{zur}
$$
   r_c^{(e)}>4.9\times{10^{-26}cm},~~~r_c^{(\mu)}>8.3\times{10^{-27}cm},$$
  $$~~~r_c^{(\tau)}>3.3\times{10^{-27}cm}\eqno(82)$$
For a scalar of $\phi^4$ theory produced at the energy scale of Eq.(80),
we estimate an upper limit for its vacuum expectation value $v$ as
$v\leq\sqrt{3m_{Pl}/\pi}$ and get an upper limit for a scalr mass as
$m_{scalar}\leq{\sqrt{3/8}m_{Pl}}$. 
These numbers give constraints for
the case of particle producation in the course of phase transitions
in the very early universe. In this sense they give the upper limit
for relic scalar particles of $\phi^4$ theory. 

Let us emphasize that the most stringent lower limits on FP
sizes as defined by de Sitter-Schwarzschild geometry, are much bigger
than the Planck length $l_{Pl}$, which justifies estimates for gravitational 
sizes of elementary particles given in the frame of classical 
general relativity.

\section{Two-lambda geometry}

In the case of non-zero value of cosmological constant at infinity
the metric has the form \cite{us97}
$$
       ds^2=\biggl(1-\frac{R_{g}(r)}{r}-\frac{\lambda r^2}{3}\biggr)dt^2
   -\biggl(1-\frac{R_{g}(r)}{r}-\frac{\lambda r^2}{3}\biggr)^{-1}dr^2$$
        $$-r^2(d\vartheta^2+\sin^2\vartheta d\varphi^2),\eqno(83)$$ 
where
$$
       R_{g}(r)=r_{g}\biggl(1-\exp{\biggl(-\frac{\Lambda r^3}
          {3r_g}\biggr)}\biggr).\eqno(84)$$
The angular components
of the stress-energy tensor are given by
$$
     T_{\vartheta}^{\vartheta}=T_{\varphi}^{\varphi}
    ={\rho}_{\Lambda}\biggl(1-\frac{\Lambda r^3}{2r_{g}}\biggr)
      \exp{\biggl(-\frac{\Lambda r^3}{3r_g}\biggr)}
          +{\kappa}^{-1}{\lambda}.\eqno(85)$$
For $r\ll (3r_g/\Lambda)^{1/3}$, the metric (83)-(84) behaves like
de Sitter metric with cosmoligical constant $\Lambda+\lambda$,
while for $r\gg (3r_g/\Lambda)^{1/3}$ it achieves the asymptotics
of the Kottlef-Trefftz solution \cite{kot,tref}
$$
       ds^2=\biggl(1-\frac{r_{g}}{r}-\frac{\lambda r^2}{3}\biggr)dt^2
        -\biggl(1-\frac{r_{g}}{r}-\frac{\lambda r^2}{3}\biggr)^{-1}dr^2$$
          $$-r^2(d\vartheta^2+\sin^2\vartheta d\varphi^2),\eqno(86)$$ 
which is frequently referred to in the literature as
the Schwarzschild-de Sitter geometry describing cosmological
black hole. Our solution (83)
represents thus the nonsingular modification of
the Kottler-Trefftz solution.
 
The quadratic invariant of the Riemann curvature tensor ${\cal R}^{2}$
$=R_{iklm}R^{iklm}$ is given by
$$
       {\cal R}^{2}=4\frac{{R_{g}}^2(r)}{r^6}
          +4\biggl(\Lambda e^{-r^3\Lambda /3r_{g}}
         -\frac{{R_{g}}(r)}{r^3}\biggr)^2$$
            $$+\biggl(\frac{2R_{g}(r)}{r^3}
     -\frac{{\Lambda}^2}{r_{g}}r^3e^{-r^3\Lambda /3r_{g}}\biggr)^2 $$
           $$~~~~~~+\frac{8{\lambda}^2}{3}+\frac{16\lambda \Lambda}{3}
         e^{-\Lambda r^3/3r_{g}}
                -\frac{4\lambda {\Lambda}^2}{3r_{g}}
                       r^3 e^{-r^3\Lambda /3r_{g}}\eqno(87)$$
For $~r\rightarrow \infty~ $, $~{\cal R}^2$  tends to the background
de Sitter curvature $~{\cal R}^{2}=8{\lambda}^2/3~$.
For $~r\rightarrow 0~$, $~{\cal R}^2$ remains finite and tends 
to the de Sitter value $~{\cal R}^{2}=8({\Lambda}+ \lambda)^2/3~$, 
which naturally appears to be the limiting value of the space-time curvature.
All other invariants of the Riemann curvature tensor are also finite.
We see that our solution is regular everywhere. 

The two-lambda spacetime has in general three horizons: a cosmological
horizon $r_{++}$, a black hole event horizon $r_{+}$ and a Cauchy
horizon $r_{-}$. The metric in general case of three horizons
is plotted in Fig.18.
\begin{figure}
\vspace{-8.0mm}
\begin{center}
\epsfig{file=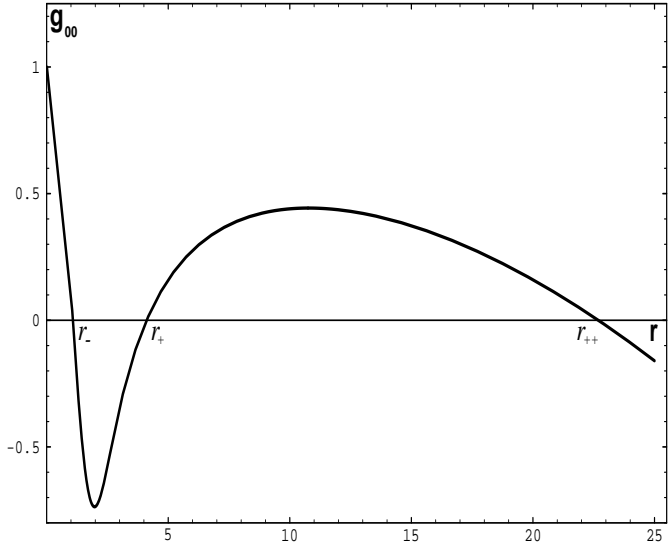,width=9.0cm,height=7.5cm}
\end{center}
\caption{
Two-lambda spherically symmetric solution  
corresponding to nonsingular cosmological black hole.
There are three horizons:
an internal horizon $~r_{-}$, a black hole horizon
$~r_{+}$, and a cosmological horizon $~r_{++}$.}
\label{fig.18}
\end{figure}        
In the region of $r_h\ll({\Lambda r_g}/{3})^{1/3}$ we can expand
the exponent in the Eq.(84) into the power series. Then we obtain
the internal horizon located at
$$
       r_{-}\simeq{\sqrt{\frac{3}{\Lambda+\lambda}}}
          \biggl[1+\frac{1}{4 r_g}\sqrt{\frac{3}{\Lambda+\lambda}}
           \biggl(\frac{\Lambda}{\Lambda+\lambda}\biggr)^2$$
        $$\biggl[1+\frac{5}{4 r_g}\sqrt{\frac{3}{\Lambda+\lambda}}
   \biggl(\frac{\Lambda}{\Lambda+\lambda}\biggr)^2\biggr]\biggr]\eqno(88)$$
for $r_g\gg{\sqrt{{3}/{\Lambda+\lambda}}
({\Lambda}/{\Lambda+\lambda})^2}$.

In the region of $r_h\gg({\Lambda r_g}/{3})^{1/3}$,
we can neglect the exponential term in Eq.(84) and find the cosmological
horizon located at
$$
    r_{++}\simeq{\sqrt{\frac{3}{\lambda}}-\frac{r_g}{2}}\eqno(89)$$
for $r_g\ll\sqrt{{3}/{\lambda}}({\Lambda}/{\lambda})$.

In the interface we are looking for a horizon in the form 
$r_h=r_g+\varepsilon, ~\varepsilon\ll r_g$. That way we get
the black hole horizon
$$
       r_{+}\simeq {r_g\biggl[1+\frac{\lambda r_g^2}{3}
      -\exp{\biggl(-\frac{\Lambda r_g^2}{3}\biggr)}\biggr]}\eqno(90)$$
for $r_g$ within the range $\sqrt{{3}/{\Lambda}}\ll r_g \ll
\sqrt{{3}/{\lambda}}$.

In general case the family of horizons can be found numerically.
Two-lambda spherically symmetric space-time has three scales of length 
$~r_{g}~$, $~r_{\Lambda}=\sqrt{3/\Lambda}~$,
and $~r_{\lambda}=\sqrt{3/\lambda}~$. Normalizing to $~r_{\Lambda}~$, 
we have two dimensionless parameters: $~M~$ (mass normalized to 
$~{G}^{-1}\sqrt{{3}/{\Lambda}}~$) 
and $~q=\sqrt{\Lambda /\lambda}~$. 
Horizons are calculated by solving the equation $g_{00}(r)=0$.
Horizon-mass diagram is plotted in Fig.19 for the case of the
density profile given by (25). 
\begin{figure}
\vspace{-8.0mm}
\begin{center}
\epsfig{file=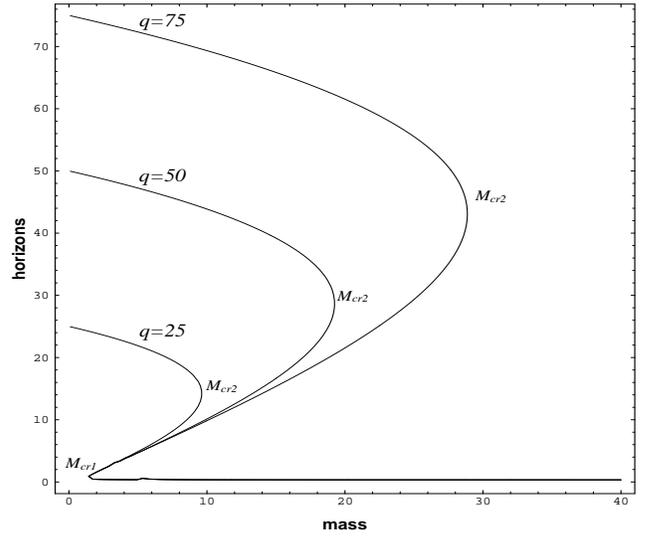,width=9.0cm,height=7.5cm}
\end{center}
\caption{
Horizon-mass diagram forn a two-lambda spacetime.
An upper limit for a BH mass $M_{cr2}$ depends on the parameter
$q={(\Lambda/\lambda)^{1/2}}$.}
\label{fig.19}
\end{figure}
Three horizons exist at the range of mass parameter 
$~M_{cr1}< M < M_{cr2}~$.
It follows that there exist two critical values of the mass $M$, 
restricting the mass of a nonsingular cosmological black hole 
from below and from above. 
A lower limit $M_{cr1}$ corresponds to the first extreme BH
state $r_{+}=r_{-}$
and is very close to the lower limit $M_{crit}$ for $\Lambda$BH
given by (27). 
An upper limit $M_{cr2}$ corresponds to the second extreme state
$r_{+}=r_{++}$ and depends on the parameter 
$q\equiv{\sqrt{{\Lambda}/{\lambda}}}$.
Depending on the mass $M$, two-lambda geometry represent five types 
of globally regular spherically symmetric configurations \cite{us98}. 
They are plotted in Fig.20 for the case of the parameter $~q~$ defining
the relation of limiting values of cosmological constant at the origin
and at infinity is given by 
$q\equiv{\sqrt{{\Lambda}/{\lambda}}}=10$. 
\begin{figure}
\vspace{-8.0mm}
\begin{center}
\epsfig{file=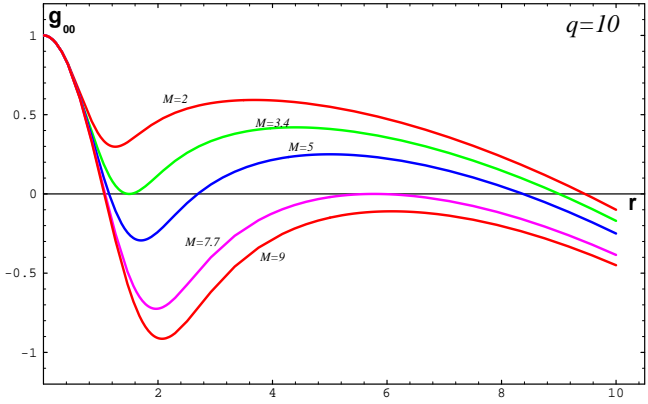,width=9.0cm,height=6.0cm} 
\end{center}
\caption{
Two-lambda configurations for the case $q=10$.
The mass $M$ is normalized to $(3/G^2 \Lambda )^{1/2}$.
Two extreme states for $\Lambda\lambda$BH are $M_{cr1}\simeq{3.4}$ 
and $M_{cr2}\simeq{7.7}$.}
\label{fig.20}
\end{figure}
Let us now specify five types of spherically symmetric configurations
described by two-lambda solution \cite{us97,us98}.
\vskip0.1in
{\bf Two-lambda black hole -}  
Within the range of masses $M_{cr1}<M<M_{cr2}$, the metric (83)
describes a nonsingular nondegenerate cosmological black hole. 
Global structure of spherically symmetric space-time
with three horizons is shown in Fig.21.
\begin{figure}
\vspace{-8.0mm}
\begin{center}
\epsfig{file=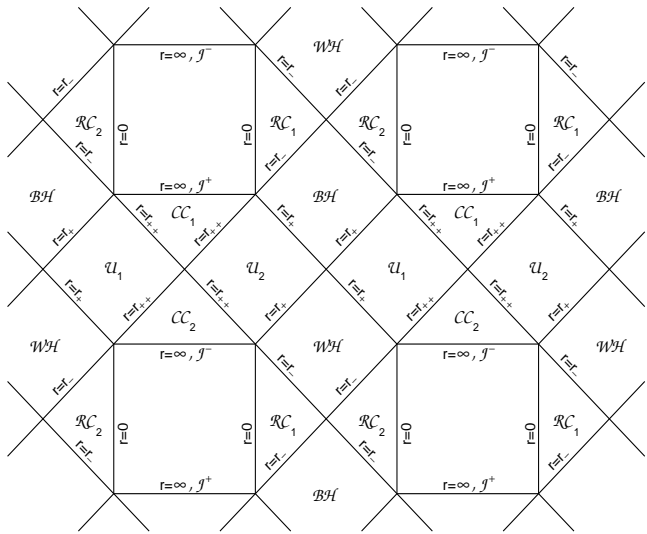,width=8.9cm,height=7.3cm}
\end{center}
\caption{
Penrose-Carter diagram for $\Lambda\lambda$ black holes.
There is an infinite sequence
of black and white holes ${\cal BH}$,${\cal WH}$, whose singularities are
replaced by future and past regular cores ${\cal RC}_1,{\cal RC}_2$, 
asymptotically de Sitter universes ${\cal U}_1,{\cal U}_2$, 
and $\lambda$ cores ${\cal CC}_1, {\cal CC}_2$
(regions beyond the cosmological horizons $r_{++}$) and spacelike
infinities.}
\label{fig.21}
\end{figure}
This diagram is similar to the case of Reissner-Nordstrom-de Sitter
geometry \cite{bron}. The essential difference is that the timelike
surface $r=0$ is regular in our case.
The global structure of space-time contains an infinite sequence 
of asymptotically de Sitter (small background $\lambda$) 
universes $~{\cal U}_{1}$, $~{\cal U}_{2}$, 
black and white holes $~{\cal BH}$, $~{\cal WH}~$ 
whose singularities are replaced with
future and past regular cores $~{\cal RC}_1$,  
${\cal RC}_2~$ (with $~\Lambda~+\lambda~$ at $r\rightarrow 0$), and
"cosmological cores" $~{\cal CC}~$ (regions between cosmological
horizons and spacelike infinities). Rectangular regions confined
by the surfaces $r=0$  and $r=\infty$ do not belong to the diagram.
 
To specify these regions, we introduce the invariant quantity
\cite{NF,victor}  
$$\Delta=g^{\mu\nu}r,_{\mu}r,_{\nu}.$$
Dependently on the sign of $\Delta$, space-time is divided into
$~R~$ and $~T~$ regions (see \cite{NF,victor}):
In the $~ R~$ regions the normal vector to the surface $r=$const,
$N_{\mu}=r,_{\mu}$ is spacelike. Therefore in the $~R~$ region
an observer on the surface $r=$const can send two radial signals:
one directed inside and the other outside of this surface.
In the $~T~$ regions the normal vector $N_{\mu}$ is timelike.
The surface $r=$const is spacelike, and both signals propagate on the 
same side of this surface. In the $~ T~$ regions any observer can cross 
the surface $r=$const only once and only in the same direction. 
The $~R~$ and $~T~$ regions are separated by horizons, where 
$\Delta=0$.
For the space-time considered here, $\Delta<0$
in the regions $~{\cal RC}$ and $~\cal U$,
and they are $~R~$ regions. 
$\Delta>0$ in the regions $~\cal{BH}$, $~\cal{WH}$ and $~\cal {CC}$,
and those are $~T~$ regions. 
 For the metric in the Kruskal form 
$~\Delta=(1/2)g_{00}^{-1}r,_u r,_v~$ \cite{victor}.
Since in the $~ T~$ regions $~\Delta>0~$, the vector $r,_u$ cannot be zero, 
and the conditions $r,_u>0$ and $r,_u<0$ are invariant. When $r,_u<0$, 
we have $~T_{-}~$ region or the region of contraction (a black hole).
When $r,_u>0$, we have an expanding $~T_{+}~$ region (a white hole).
Near horizons the sign of $~T~$  tells us that the vector
$r,_u$ enters or goes out of the $~R~$ region.
\vskip0.1in
{\bf The lower mass extreme state -} The critical value of mass  
$M_{cr1}$, at which
the internal horizon $r_{-}$ coincides with the black hole horizon $r_{+}$,
corresponds to the first extreme black hole state. For $M<M_{cr1}$
there is no black hole (see Fig.18), so $M_{cr1}$
puts the lower limit for a black hole mass. 
It is given by (27) 
and practically does not depend on the parameter $q=\sqrt{\Lambda/\lambda}$.  
This extreme black hole is shown in Fig.19 for the dimensionless
value of mass $M=3.4$.

Let us compare the situation with the Schwarzschild-de Sitter
family of singular black holes. Those black holes 
have masses between zero and the size of the cosmological horizon 
(see, e.g.,\cite{bousso}). Replacing a singularity
with a cosmological constant $\Lambda$ results in appearance
of the lower limit for a mass of cosmological black hole. It leads 
thus to the existence of the new type spherically symmetric configuration 
- the extreme neutral nonsingular cosmological black hole 
whose internal horizon coincides with a black hole horizon.
\vskip0.1in
 {\bf The upper mass extreme state -} The upper limit $M_{cr2}$,
at which  the black hole horizon $r_{+}$ coincides with the cosmological
horizon $r_{++}$, corresponds to the degenerate black hole 
existing also in the Schwarzschild-de Sitter family  
and known as the Nariai solution \cite{nariai}. 
What we found here for the case of $M=M_{cr2}$ is the nonsingular
modification of the Nariai solution. 
As by-product of replacing a singularity by a de Sitter-like core
we have in this case additional internal horizon (see Fig.18).
The value of $~M_{cr2}~$ depends essentially on the parameter 
$q=\sqrt{\Lambda/\lambda}$ (see Fig.18).
The second extreme black hole is shown in Fig.19 for $M=7.7$.
\vskip0.1in
{\bf Soliton-like configurations -}
Beyond the limiting masses $M_{cr1}$ and $M_{cr2}$, there exist two
different types of nonsingular spherically symmetric configurations:

(i) In the range of mass parameter $M<M_{cr1}$ two-lambda solution (83) 
describes spherically symmetric selfgravitating particlelike structure 
at the de Sitter background 
($\Lambda_{\mu\nu}\rightarrow {\lambda g_{\mu\nu}}$ as 
$r\rightarrow{\infty}$). This case differs 
from the case of selfgravitating particlelike structure at the flat space 
background Fig.2 by existence of the cosmological 
horizon (see Fig.19 with $M=2$).

(ii) For $M>M_{cr2}$ we have quite different type of nonsingular one-horizon 
spherically symmetric configuration. It differs essentially from the 
Schwarzschild-de Sitter case by existence of an internal horizon 
(see Fig.19 with $M=9$),
which comes from replacing a singularity by a de Sitter regular core.
It is rather new type of configuration which can be called  
"de Sitter bag".

\section{Summary and discussion}

We have presented here the nonsingular modification of 
the Schwarzschild-de Sitter family of black hole solutions, 
obtained by replacing a singularity with a
de Sitter regular core
with $\Lambda$ at the scale of symmetry restoration.
Main features of these geometries: 

1) algebraic structure of a stress-energy tensor as describing
spherically symmetric vacuum corresponding to $r-$dependent
cosmological term $\Lambda_{\mu\nu}$

2) de Sitter value for spacetime curvature at $r=0$

3) global structure of space-time,

4) existence of the lower limit for a black hole mass, 

5) second-order phase transition in the course of Hawking evaporation 
and zero final temperature -

result from the boundary conditions and 
from the condition $g_{00}=-g_{11}^{-1}$ imposed for a metric. 
They do not depend on a particular form of the density 
profile $\Lambda_t^t(r)$ if it satisfies condition of finiteness 
of mass and density.

Global structure of space-time in the case of
vacuum nonsingular black holes contains an infinite sequence of
black and white holes whose singularities are
replaced with future and past regular cores, at the background 
of Minkowski or de Sitter space.

$\Lambda$ white hole corresponds to nonsingular vacuum-dominated 
cosmology governed by the time-dependent cosmological term 
$\Lambda_{\mu\nu}$ \cite{us2000}.
It models the initial stages of the Universe evolution - 
nonsingular nonsimultaneous big bang followed
by anisotropic Kasner-type stage at which most of a mass 
is produced \cite{us99}.

Instability of de Sitter vacuum at the origin can lead to
creation of baby universes inside of vacuum black holes.
The probability of a single tunneling event is
very small, but in our case there exists an infinite number 
of white hole structures inside each black hole.
This highly magnifies the probability of birth of a baby universe
via quantum tunneling in one of them. 

Replacing a central singularity with a value of a cosmological
constant, resulted in
appearance of new type of globally regular vacuum-dominated
configurations described by the spherically symmetric solutions 
- vacuum self-gravitating particlelike structures at the background
of Minkowski or de Sitter space.

Following Schwarzschild, we can say that "it is always nice
to have analytic solution of a simple form".
It allows us to investigate behaviour qualitatively
and gives a chance to reveal features which otherwise could escape 
from analysis. For example, numerical analysis of non-Abelian black 
holes revealed existence, among them, a "neutral" type for which 
a non-Abelian structure can be approximated as a vacuum
core of uniform density \cite{maeda}
(suggesting existence of de Sitter-Schwarzschild black hole), 
however two essential features of a black hole of this type - 
existence of the lower limit for a black hole mass $M_{crit}$
and vanishing the temperature  in the course of evaporation
- remained concealed.  
Analytic solution tells us that
the form of a temperature-mass diagram is determined by the
existence of two horizons. Temperature drops to zero and evaporation stops 
when the horizons merge in the course of evaporation.

These results allow us to speculate about possible end point 
for evaporation of a vacuum black hole.
One possibility is connected with back-reaction effects. 
If the back-reaction effects would diminish a mass of the extreme
configuration, then a black hole could disappear leaving 
a cold remnant - a recovered particlelike structure - in its place.  
We are currently working on investigating its stability.
Other possibility is suggested by the global structure of the extreme 
black hole. 
In this case the Killing vector $K$ is not spacelike up to the
cosmological horizon, and in principle 
nothing could prevent internal de Sitter core from exploding. 
If it would occur, a black hole would evaporate completely.
Let us note, that in any case information about baby universes
inside a black hole is lost in the course of evaporation.
\vskip0.2in
{\bf Acknowledgement}

This talk was supported by the University of Warmia and Mazury
in Olsztyn.

\end{document}